\documentclass{amsart}
\usepackage{amssymb}
\usepackage{amsfonts}
\usepackage{amsmath}
\usepackage{graphicx,geometry}
\usepackage{bm,cite}

\newtheorem{thm}{Theorem}

\newtheorem{lem}{Lemma}

\begin{document}

\title{Extended Krein-Adler theorem for the translationally shape invariant
potentials}
\author{David G\'omez-Ullate}\thanks{This work has been partially supported by the Spanish MINECO-FEDER Grants MTM2009-06973, MTM2012-31714, the Catalan Grant
2009SGR--859 and the Canadian NSERC grant RGPIN-228057-2009.} 
\address{ Departamento de F\'isica Te\'orica II, Universidad Complutense de
Madrid, 28040 Madrid, Spain.}
\author{ Yves Grandati}
\address{Equipe BioPhysStat, LCP A2MC, Universit\'e de Lorraine--Site de Metz,
1 Bvd D. F. Arago, F-57070, Metz, France.}
\author{Robert Milson}
\address{Department of Mathematics and Statistics, Dalhousie University,
Halifax, NS, B3H 3J5, Canada.}

\begin{abstract}
Considering successive extensions of primary translationally shape invariant
potentials, we enlarge the Krein-Adler theorem to mixed chains of state
adding and state-deleting Darboux-B\"{a}cklund transformations.
It allows us to establish novel bilinear Wronskian and determinantal
identities for classical orthogonal polynomials.
\end{abstract}

\maketitle

\section{\protect\bigskip Introduction}

In quantum mechanics the construction of solvable potentials is a question
of central interest and a subject of long time studies. The most efficient
way to produce new systems which are exactly solvable in closed analytical
form, starting from known ones is certainly the Darboux or Darboux-B\"{a}%
cklund transformation (DBT) \cite{darboux}. Using one given eigenfunction of
the initial potential (the seed function of the transformation), the DBT
associates to this last a new potential, called an extension, which is
formally "isospectral" to the previous one. It is then possible to enchain
the DBT to generate chains of such isospectral extensions and in 1955, Crum 
\cite{crum} has proven that for non-degenerate chains, both the extended
potentials and their eigenfunctions can be written in terms of Wronskians of
eigenfunctions of the initial potential (the seed functions of the chain).

To go beyond the formal level and define equivalent boundary value problems
for the systems related by a chain of DBT, it is necessary to ensure the
regular character of the extended potentials. Crum \cite{crum} has shown
that this is achieved if we use a complete chain, i.e. a set of seed functions
which are consecutive eigenstates starting from the fundamental one. The
obtained extensions corresponds exactly to the hierarchy of superpartner
potentials obtained in SUSY quantum mechanics \cite{cooper,Dutt}. A more
general answer has been provided shortly after by Krein \cite{krein} and
retrieved later by Adler \cite{adler}. The so called Krein-Adler (KA) chains
are also based on eigenfunctions of the initial potential but are lacunary
with specific even gaps. The successive DBT in a KA chain lead to extensions
in the spectrum of which some levels have been suppressed and for this
reason they are said to be "state-deleting".

There is only a handful of classically known potentials for which it is to
be possible to determine explicitly all the bound states in terms of
elementary transcendental functions and to write explicitly their energies
in terms of the quantum number. They all belong to category of
translationally shape invariant potentials (TSIP) \cite%
{cooper,Dutt,gendenshtein,grandati} and they possess the characteristic
feature to have bound states which are (up to a gauge factor) classical
orthogonal polynomials in an adapted variable. Starting from these TSIP, the
extensions generated by a \ KA chain are then rational in this variable.

During the four last years notable progress has been made in the study of
rational extensions of TSIP and of their intimate connection with the
recently discovered \textit{exceptional orthogonal polynomials} \cite%
{gomez,gomez2,gomez3,gomez4,gomez5,gomez6,gomez7,gomez8,quesne1,quesne,quesne2,quesne3,quesne4,quesne5,quesne6,quesne7,odake,sasaki,ho,ho2,ho3,odake2,odake3,odake5,sasaki2,dutta,grandati2,grandati3,grandati4,grandati5,grandati6,grandati7,grandati8,ramos1,ramos2}%
.

A systematic procedure generating such extensions has been proposed \cite%
{grandati2,grandati3,grandati4,grandati5,grandati6,grandati7,grandati8}. It
rests on the use of DBT or chains of DBT based on unphysical seed functions
belonging to the disconjugacy sector of the initial equation, which are
obtained from the eigenstates by a regularization procedure and which then
share the quasi-polynomial character of the physical states. The
regularization is achieved by the use of specific discrete symmetries of the
original potential or, for finite bound state systems, by prolongation of
the quantum number out of the physical domain of values. It allowed in
particular to show the existence of infinite towers of strictly isospectral
extensions which share with the original potential (then called a primary
TSIP or PTSIP) the same shape invariance properties and constitute novel
families of secondary TSIP. It also lead to the discovery of an enlarged
shape invariance property satisfied by some families of extensions of
potential with finite bound state spectrum \cite{quesne6,grandati6}. The
eigenstates of all these families of extended potentials are expressible in
terms of various types of exceptional orthogonal polynomials \cite%
{gomez,gomez2,gomez3,gomez4,gomez5,gomez6,gomez7,gomez8}.

It appears that all the PTSIP possess at least one common discrete symmetry,
called type 3 symmetry, which corresponds to a simultaneous change of the
sign of all the parameters on which  the potential depends. As opposed to
the other possible symmetries (type 1 and 2) the DBT obtained by using this
symmetry as a regularization procedure can generate only quasi-isospectral (or
essentially isospectral) extensions in which spectrum contains one extra
level (state-adding DBT).  The position of this extra level corresponds to a
negative integer value of the spectral parameter.

In this paper we consider extensions of PTSIP built from mixed chains, i.e.
containing state-adding and state-deleting DBT. The specific case of the
harmonic oscillator, where the type 3 symmetry corresponds in fact to a
"Wick rotation" of the independent variable, has been the subject of long
time investigation \cite{shnol',bagrov,bagrov1,fernandez,tkachuk}. Oblomkov
has in particular shown that the extensions obtained from general mixed
chains coincide with the rational extensions of the harmonic oscillator
which possess the trivial monodromy property \cite{oblomkov}. In a recent
work we have proven that this set of extensions contains those which are
associated to hamiltonians exactly solvable by polynomials and the spectrum
of which being subtended by systems of exceptional Hermite polynomials \cite%
{ggm}, providing a precise description of the properties of these $X$-Hermite polynomials.

We show here that every PTSIP possesses a "reverse shape invariance property"
which allows to enlarge the Krein-Adler theorem to some mixed chains. This
provides in particular new bilinear determinantal or Wronskians identities
for the classical orthogonal polynomials. We would like to mention that
during the writing of the present article, parallel results have been presented in \cite%
{odake4}, although from a different point of view, 

In the first two sections we give basic facts concerning DBT and
non-degenerate chains of DBT. Next we give new explicit proofs of Wronskian
formulas for the extended potentials and their eigenfunctions in the case of
degenerate chains. In the following parts, we recall the regularity
criteria for DBT and chains of DBT, in particular the Krein-Adler theorem.
We then describe the essential features of TSIP and the way to use specific
symmetries of PTSIP to generate their rational extensions. Using seed
functions obtained via the type 3 symmetry which generate state-adding DBT,
we establish the reverse shape invariance condition associated to complete
chains of such state-adding DBT. It allows to prove an extended Krein-Adler
theorem for mixed chains and combining it to the general Wronskian formulas
previously established, we give new bilinear Wronskian formulas for the
eigenfunctions of the PTSIP. Finally we apply these results to specific
examples (harmonic and isotonic oscillators, trigonometric Darboux-P\"{o}%
schl-Teller potential, Morse and Kepler-Coulomb potentials) showing how it
allows to obtain new bilinear Wronskians or determinantal identities for
Hermite, Laguerre and Jacobi polynomials.

\section{Formal Darboux-B\"{a}cklund Transformations (DBT) and Crum formulas}

We consider a one dimensional Hamiltonian $\widehat{H}=-d^{2}/dx^{2}+V(x),\
x\in I\subset \mathbb{R}$ and the associated Schr\"{o}dinger equation%
\begin{equation}
\psi _{\lambda }^{\prime \prime }(x)+\left( E_{\lambda }-V(x)\right) \psi
_{\lambda }(x)=0,  \label{EdS}
\end{equation}%
$\psi _{\lambda }(x)$ being a formal eigenfunction of $\widehat{H}$ for the
eigenvalue $E_{\lambda }$. In the following we suppose that, with Dirichlet
boundary conditions on $I$, $\widehat{H}$ admits a discrete spectrum of
energies and eigenstates of the $\left( E_{n},\psi _{n}\right) _{n\in
\left\{ 0,...,n_{\max }\right\} \mathbb{\subseteq N}}$ where, without loss
of generality, we can always suppose that the ground level of $\widehat{H}$
is zero ($E_{0}=0$).

The \textit{Riccati-Schr\"{o}dinger (RS)} function $w_{\lambda }(x)=-\psi
_{\lambda }^{\prime }(x)/\psi _{\lambda }(x)$ associated to $\psi _{\lambda
} $ satisfies the corresponding \textit{Riccati-Schr\"{o}dinger (RS)}
equation \cite{grandati}

\begin{equation}
-w_{\lambda }^{\prime }(x)+w_{\lambda }^{2}(x)=V(x)-E_{\lambda }.
\label{edr4}
\end{equation}

If the set of general Riccati equations is invariant under the group $%
\mathcal{G}$ of smooth $SL(2,\mathbb{R})$-valued curves $Map(\mathbb{R},SL(2,%
\mathbb{R}))$ \cite{carinena2,Ramos}, the particular subclass of Riccati-Schr%
\"{o}dinger equations is preserved by a specific subset of $\mathcal{G}$,
called the \textit{Darboux-B\"{a}cklund Transformations (DBT)}. Such
transformation can be built from any solution $w_{\nu }(x)$ of the initial
RS equation Eq(\ref{edr4}) as \cite{carinena2,Ramos,grandati}

\begin{equation}
w_{\lambda }(x)\overset{A\left( w_{\nu }\right) }{\rightarrow }w_{\lambda
}^{\left( \nu \right) }(x)=-w_{\nu }(x)+\frac{E_{\lambda }-E_{\nu }}{w_{\nu
}(x)-w_{\lambda }(x)},  \label{transfoback2}
\end{equation}%
where $\lambda \neq \nu $.

The function $w_{\lambda }^{\left( \nu \right) }$ is then a solution of the RS equation

\begin{equation}
-w_{\lambda }^{\left( \nu \right) \prime }(x)+\left( w_{\lambda }^{(\nu
)}(x)\right) ^{2}=V^{\left( \nu \right) }(x)-E_{\lambda },
\label{eqtransform}
\end{equation}%
with the same energy $E_{\lambda }$ as in Eq(\ref{edr4}) but with a modified
potential

\begin{equation}
V^{\left( \nu \right) }(x)=V(x)+2w_{\nu }^{\prime }(x),  \label{pottrans}
\end{equation}%
that we call an \textit{extension} of $V(x)$.

To $w_{\lambda }^{\left( \nu \right) }$ corresponds an eigenfunction $\psi
_{\lambda }^{\left( \nu \right) }=\exp \left( -\int dxw_{\lambda }^{(\nu
)}(x)\right) $ of the extended hamiltonian $\widehat{H}^{\left( \nu \right)
}=-d^{2}/dx^{2}+V^{\left( \nu \right) }(x)$, which can be written

\begin{equation}
\psi _{\lambda }^{\left( \nu \right) }(x)\sim \widehat{A}\left( w_{\nu
}\right) \psi _{\lambda }(x),  \label{foDBT}
\end{equation}%
where $\widehat{A}\left( w_{\nu }\right) $ is a first order operator given by

\begin{equation}
\widehat{A}\left( w_{\nu }\right) =d/dx+w_{\nu }(x).  \label{opA}
\end{equation}

W swee that (\ref{foDBT}) can then be rewritten as the \textit{Darboux-Crum formula}

\begin{equation}
\psi _{\lambda }^{\left( \nu \right) }(x)\sim \frac{W\left( \psi _{\nu
},\psi _{\lambda }\mid x\right) }{\psi _{\nu }(x)},  \label{foDBTwronsk}
\end{equation}%
where $W\left( y_{1},...,y_{m}\mid x\right) $ denotes the Wronskian of the
family of functions $y_{1},...,y_{m}$

\begin{equation}
W\left( y_{1},...,y_{m}\mid x\right) =\left\vert 
\begin{array}{ccc}
y_{1}\left( x\right) & ... & y_{m}\left( x\right) \\ 
... &  & ... \\ 
y_{1}^{\left( m-1\right) }\left( x\right) & ... & y_{m}^{\left( m-1\right)
}\left( x\right)%
\end{array}%
\right\vert .  \label{wronskien}
\end{equation}

The eigenfunction $\psi _{\nu }$ is called the \textit{seed function} of the
DBT $A(w_{\nu })$ and $V^{\left( \nu \right) }$ and $\psi _{\lambda
}^{\left( \nu \right) }$ are the \textit{Darboux transforms} of $V$ and $%
\psi _{\lambda }$ respectively.

Note that $A(w_{\nu })$ annihilates $\psi _{\nu }$ and consequently the
formulas Eq(\ref{foDBT}) and Eq(\ref{foDBTwronsk}) allow to obtain an
eigenfunction of $V^{\left( \nu \right) }$ for the eigenvalue $E_{\lambda }$
only when $\lambda \neq \nu $. Nevertheless, we can readily verify that $%
1/\psi _{\nu }(x)$ is such an eigenfunction. By extension, we then define
the "image" by $A(w_{\nu })$ of the seed eigenfunction $\psi _{\nu }$ itself
as 
\begin{equation}
\psi _{\nu }^{\left( \nu \right) }(x)\sim \frac{1}{\psi _{\nu }(x)}.
\label{psinunu}
\end{equation}

\section{Formal chains of Darboux-B\"acklund Transformations}

At the formal level, the DBT can be straightforwardly iterated and a \textit{%
chain of }$m$\textit{\ DBT} can be simply described by the following scheme

\begin{equation}
\left\{ 
\begin{array}{c}
\psi _{\lambda }\overset{A(w_{\nu _{1}})}{\rightarrowtail }\psi _{\lambda
}^{\left( \nu _{1}\right) }\overset{A(w_{\nu _{2}}^{\left( N_{1}\right) })}{%
\rightarrowtail }\psi _{\lambda }^{\left( N_{2}\right) }...\overset{A(w_{\nu
_{m}}^{\left( N_{m-1}\right) })}{\rightarrowtail }\psi _{\lambda }^{\left(
N_{m}\right) } \\ 
V\overset{A(w_{\nu _{1}})}{\rightarrowtail }V^{\left( \nu _{1}\right) }%
\overset{A(w_{\nu _{2}}^{\left( N_{1}\right) })}{\rightarrowtail }V^{\left(
N_{2}\right) }...\overset{A(w_{\nu _{m}}^{\left( N_{m-1}\right) })}{%
\rightarrowtail }V^{\left( N_{m}\right) },%
\end{array}%
\right.  \label{diagn}
\end{equation}%
where $N_{j}$ denotes the $j$-uple $\left( \nu _{1},,...,\nu _{j}\right) $
(with $N_{1}=\nu _{1}$) which completely characterizes the chain. We note $%
\left( N_{m},\nu _{m+1},...,\nu _{m+k}\right) $ the chain obtained by adding
to the chain $N_{m}$ the DBT associated to the successive eigenfunctions $%
\psi _{\nu _{m+1}}^{\left( N_{m}\right) },...,\psi _{\nu _{m+k}}^{\left(
N_{m+k-1}\right) }$.

$\psi _{\lambda }^{\left( N_{m}\right) }$ is an eigenfunction associated to
the eigenvalue $E_{\lambda }$ of the potential (see Eq(\ref{pottrans}))

\begin{equation}
V^{\left( N_{m}\right) }(x)=V(x)+2\sum_{j=1}^{m}\left( w_{\nu _{j}}^{\left(
N_{j-1}\right) }(x)\right) ^{\prime }=V(x)-2\sum_{j=1}^{m}\left( \log \left(
\psi _{\nu _{j}}^{\left( N_{j-1}\right) }(x)\right) \right) ^{\prime \prime }
\label{potnstep}
\end{equation}%
and can be written as (cf Eq(\ref{foDBT}) and Eq(\ref{foDBTwronsk}))

\begin{equation}
\psi _{\lambda }^{\left( N_{m}\right) }(x)=\widehat{A}\left( w_{\nu
_{m}}^{\left( N_{m-1}\right) }\right) \psi _{\lambda }^{\left(
N_{m-1}\right) }(x)=\widehat{A}\left( w_{\nu _{m}}^{\left( N_{m-1}\right)
}\right) ...\widehat{A}\left( w_{\nu _{1}}\right) \psi _{\lambda }(x),
\label{etats n}
\end{equation}%
that is,

\begin{equation}
\psi _{\lambda }^{\left( N_{m}\right) }(x)=\frac{W\left( \psi _{\nu
_{m}}^{\left( N_{m-1}\right) },\psi _{\lambda }^{\left( N_{m-1}\right) }\mid
x\right) }{\psi _{\nu _{m}}^{\left( N_{m-1}\right) }(x)}.  \label{fon}
\end{equation}

A chain is \textit{non-degenerate} if all the spectral indices $\nu _{i}$ of
the chain $N_{m}$ are distinct and is \textit{degenerate} if some of them
are repeated in the chain. For non-degenerate chains Crum has derived very
useful formulas for the extended potentials and their eigenfunctions in
terms of Wronskians of eigenfunctions of the initial potential \cite%
{crum,matveev}.

\subsection*{Crum's formulas}

When all the $\nu _{j}$ and $\lambda $ are
distinct, we have

\begin{equation}
\psi _{\lambda }^{\left( N_{m}\right) }(x)=\frac{W^{\left( N_{m},\lambda
\right) }\left( x\right) }{W^{\left( N_{m}\right) }\left( x\right) }
\label{etats n3}
\end{equation}%
\textit{and}

\begin{equation}
V^{\left( N_{m}\right) }(x)=V(x)-2\left( \log W^{\left( N_{m}\right) }\left(
x\right) \right) ^{\prime \prime },  \label{potnstep2}
\end{equation}%
where $W^{\left( N_{m}\right) }\left( x\right) =W\left( \psi _{\nu
_{1}},...,\psi _{\nu _{m}}\mid x\right) $\textit{.}

A very direct proof of this result can be obtained by using the Jacobi
identity for Wronskians (see \cite{vein} for an extensive discussion of this result).

\subsection*{Wronskian Jacobi formulas}

Let $A=\left[ a_{ij}\right] $ be a $n\times n$
matrix. We denote by $A_{1...n}^{1...n}$ the determinant of $A$
 and $A_{q_{1}...q_{r}}^{p_{1}...p_{r}}$ the retainer minor of
order $r$ which is the subdeterminant obtained by
deleting the elements belonging to the lines $\left\{ 1,...,n\right\}
/\left\{ p_{1}...p_{r}\right\} $ and to the columns $\left\{
1,...,n\right\} /\left\{ q_{1}...q_{r}\right\} $ and retaining all
other elements. The \textit{Jacobi identity} reads

\begin{equation}
A_{1...n}^{1...n}A_{1...n-2}^{1...n-2}-A_{1...n-1}^{1...n-1}A_{1...n-2,n}^{1...n-2,n}+A_{1...n-2,n}^{1...n-2,n-1}A_{1...n-2,n-1}^{1...n-2,n}=0.
\label{jac}
\end{equation}

This identity leads to the following \textit{Wronskian Jacobi formula
}
\begin{equation}
W\left( y_{1},...,y_{n}\mid x\right) =\frac{W\left( W\left(
y_{1},...,y_{n-1}\right) ,W\left( y_{1},...,y_{n-2},y_{n}\right) \mid
x\right) }{W\left( y_{1},...,y_{n-2}\mid x\right) }.  \label{wronkjac}
\end{equation}

\textit{Proof:} The proof is a direct consequence of the Jacobi identity if we take $%
A_{1...n}^{1...n}=W\left( y_{1},...,y_{n}\mid x\right) $ and we note that

\begin{equation}
A_{1...n-2,n}^{1...n-2,n}=\left\vert 
\begin{array}{cccc}
y_{1} & ... & y_{n-2} & y_{n} \\ 
... &  & ... & ... \\ 
y_{1}^{\left( n-2\right) } & ... & y_{n-2}^{\left( n-2\right) } & 
y_{n}^{\left( n-2\right) } \\ 
y_{1}^{\left( n\right) } & ... & y_{n-2}^{\left( n\right) } & y_{n}^{\left(
n\right) }%
\end{array}%
\right\vert =W^{\prime }\left( y_{1},...,y_{n-2},y_{n}\mid x\right),
\end{equation}%
and in the same manner

\begin{equation}
A_{1...n-2,n}^{1...n-2,n-1}=W^{\prime }\left( y_{1},...,y_{n-1}\right) .
\end{equation}%

\textit{Proof of the Crum formulas:} Using in the Wronskian Jacobi identity
the following property of Wronskians \cite{muir}

\begin{equation}
W\left( uy_{1},...,uy_{m}\mid x\right) =u^{m}W\left( y_{1},...,y_{m}\mid
x\right) ,  \label{factwronsk}
\end{equation}%
this gives

\begin{equation}
\frac{W\left( y_{1},...,y_{m},y\mid x\right) }{W\left( y_{1},...,y_{m}\mid
x\right) }=\frac{W\left( \frac{W\left( y_{1},...,y_{m}\right) }{W\left(
y_{1},...,y_{m-1}\right) },\frac{W\left( y_{1},...,y_{m-1},y\right) }{%
W\left( y_{1},...,y_{m-1}\right) }\mid x\right) }{\frac{W\left(
y_{1},...,y_{m}\mid x\right) }{W\left( y_{1},...,y_{m-1}\mid x\right) }}
\end{equation}%
and comparing to Eq(\ref{fon}), we obtain the Crum formula for the
eigenfunctions Eq(\ref{etats n3}). Inserting this result in Eq(\ref{potnstep}%
), we then deduce the Crum formula for the potential Eq(\ref{potnstep2}). 

The eigenfunctions $\psi _{\nu _{1}},...,\psi _{\nu _{m}}$ of $V$ are called
the \textit{seed functions} of the chain of DBT associated to the $m$-uple
 of spectral indices $N_{m}=\left( \nu _{1},...,\nu _{m}\right) $.

\section{Wronskian formulas for the images of the seed functions.}

Being obtained by recurrence starting from the one step Wronskian formula Eq(%
\ref{foDBTwronsk}) which is not relevant when $\lambda =\nu $, the Crum
formulas are restricted to the values of $\lambda \notin N_{m}$.
Nevertheless, we can also obtain Wronskian formulas for the image by a chain
of DBT of one of the seed function associated to the chain. First note that,
to obtain $\psi _{\nu _{k}}^{\left( N_{m}\right) }$, $1\leq k<m$, we can
make use the Crum formulas if we start only from the $\left( k+1\right)
^{th} $ step. By using the eigenfunctions $\left( \psi _{\nu _{k+1}}^{\left(
N_{k}\right) },...,\psi _{\nu _{m}}^{\left( N_{k}\right) }\right) $ of $%
V^{\left( N_{k}\right) }$, $N_{k}=\left( \nu _{1},...,\nu _{k}\right) $, as
seed functions and apply the corresponding chain of $m-k$ DBT to (see Eq(%
\ref{psinunu}))%
\begin{equation}
\psi _{\nu _{k}}^{\left( N_{k}\right) }\sim \frac{1}{\psi _{\nu
_{k}}^{\left( N_{k-1}\right) }}=\frac{W^{\left( N_{k-1}\right) }\left(
x\right) }{W^{\left( N_{k}\right) }\left( x\right) },
\end{equation}%
we obtain

\begin{equation}
\psi _{\nu _{k}}^{\left( N_{m}\right) }(x)=\frac{W\left( \psi _{\nu
_{k+1}}^{\left( N_{k}\right) },...,\psi _{\nu _{m}}^{\left( N_{k}\right)
},\psi _{\nu _{k}}^{\left( N_{k}\right) }\mid x\right) }{W\left( \psi _{\nu
_{k+1}}^{\left( N_{k}\right) },...,\psi _{\nu _{m}}^{\left( N_{k}\right)
}\mid x\right) }.
\end{equation}

\begin{lem}\label{lem1}

Consider the formal non-degenerate chain of DBT characterized by the 
$m$-uple of spectral indices $N_{m}=\left( \nu _{1},...,\nu
_{m}\right)$. If $\left\{ \nu _{j_{1}},...,\nu _{j_{l}}\right\}$
is a subset of $\left\{ \nu _{1},...,\nu _{m}\right\} $, 
$N_{m}/\nu _{j_{1}},...,\nu _{j_{l}}$ means the \mbox{$\left( m-l\right)$-uple} obtained by suppressing into $\left( \nu _{1},...,\nu
_{m}\right) $ the elements $\left( \nu _{j_{1}},...,\nu
_{j_{l}}\right) $. Then, for any $1\leq k\leq m$ we have

\begin{equation}
\psi _{\nu _{k}}^{\left( N_{m}\right) }(x)\sim \frac{W^{\left( N_{m}/\nu
_{k}\right) }\left( x\right) }{W^{\left( N_{m}\right) }\left( x\right) },
\label{imageseed}
\end{equation}%
where the ordering of the indices in $N_{m}$ is arbitrary.
\end{lem}
\textit{Proof:}
In the particular case $k=m$, taking account of Eq(\ref{psinunu}) and Eq(\ref%
{etats n3}), the result is immediate

\begin{equation}
\psi _{\nu _{m}}^{\left( N_{m}\right) }(x)\sim \frac{1}{\psi _{\nu
_{m}}^{\left( N_{m-1}\right) }(x)}\sim \frac{W^{\left( N_{m}/\nu _{m}\right)
}\left( x\right) }{W^{\left( N_{m}\right) }\left( x\right) }.  \label{24}
\end{equation}

It results that if the ordering of the successive DBTs in the chain is, as
in the case of the usual Crum formulas, unimportant (up to a constant
factor) Eq(\ref{24}) implies Eq(\ref{imageseed}). We now proceed by
induction to show that is indeed the case.

For $m=2$, $N_{2}=\left( \nu _{1},\nu _{2}\right) $, using standard
properties of the Wronskians \cite{muir}, Eq(\ref{foDBTwronsk}) and Eq(\ref%
{psinunu}), we deduce

\begin{equation}
\psi _{\nu _{1}}^{\left( N_{2}\right) }(x)=\frac{W\left( \psi _{\nu
_{1}}^{\left( \nu _{1}\right) },\psi _{\nu _{2}}^{\left( \nu _{1}\right)
}\mid x\right) }{\psi _{\nu _{2}}^{\left( \nu _{1}\right) }(x)}\sim \frac{%
W\left( 1,W\left( \psi _{\nu _{1}},\psi _{\nu _{2}}\right) \mid x\right) }{%
\psi _{\nu _{1}}(x)W\left( \psi _{\nu _{1}},\psi _{\nu _{2}}\mid x\right) }%
\sim \frac{W^{\prime }\left( \psi _{\nu _{1}},\psi _{\nu _{2}}\mid x\right) 
}{\psi _{\nu _{1}}(x)W\left( \psi _{\nu _{1}},\psi _{\nu _{2}}\mid x\right) }%
.  \label{25}
\end{equation}

With the help of the Wronskian theorem \cite{messiah}

\begin{equation}
W^{\prime }\left( \psi _{\nu _{1}},\psi _{\nu _{2}}\mid x\right) =\left(
E_{\nu _{1}}-E_{\nu _{2}}\right) \psi _{\nu _{1}}(x)\psi _{\nu _{2}}(x),
\end{equation}%
Eq(\ref{25}) can be rewritten as

\begin{equation}
\psi _{\nu _{1}}^{\left( N_{2}\right) }(x)\sim \frac{\psi _{\nu _{2}}(x)}{%
W\left( \psi _{\nu _{1}},\psi _{\nu _{2}}\mid x\right) }\sim 1/\psi _{\nu
_{1}}^{\left( \nu _{2}\right) }(x)=\psi _{\nu _{1}}^{\left( \nu _{2},\nu
_{1}\right) }(x).
\end{equation}

Suppose then Eq(\ref{imageseed}) is true for a $m$-step DBT. Consider then a 
$\left( m+1\right) $-step chain. Since at the preceding step, the ordering
is \ not important, we can write

\begin{equation}
\psi _{\nu _{k}}^{\left( N_{m+1}\right) }(x)=\frac{W\left( \psi _{\nu
_{k}}^{\left( N_{m}\right) },\psi _{\nu _{m+1}}^{\left( N_{m}\right) }\mid
x\right) }{\psi _{\nu _{m+1}}^{\left( N_{m}\right) }(x)}=\frac{W\left( \psi
_{\nu _{k}}^{\left( \nu _{1},...,\nu _{k-1},\nu _{k+1},...,\nu _{m},\nu
_{k}\right) },\psi _{\nu _{m+1}}^{\left( \nu _{1},...,\nu _{k-1},\nu
_{k+1},...,\nu _{m},\nu _{k}\right) }\mid x\right) }{\psi _{\nu
_{m+1}}^{\left( \nu _{1},...,\nu _{k-1},\nu _{k+1},...,\nu _{m},\nu
_{k}\right) }(x)}.
\end{equation}

This can be rewritten

\begin{eqnarray}
\psi _{\nu _{k}}^{\left( N_{m+1}\right) }(x) &=&\frac{W\left( W\left( \psi
_{\nu _{m+1}}^{\left( N_{m}/\nu _{k}\right) },\psi _{\nu _{k}}^{\left(
N_{m}/\nu _{k}\right) }\right) /\psi _{\nu _{k}}^{\left( N_{m}/\nu
_{k}\right) },1/\psi _{\nu _{k}}^{\left( N_{m}/\nu _{k}\right) }\mid
x\right) }{W\left( \psi _{\nu _{m+1}}^{\left( N_{m}/\nu _{k}\right) },\psi
_{\nu _{k}}^{\left( N_{m}/\nu _{k}\right) }\right) /\psi _{\nu _{k}}^{\left(
N_{m}/\nu _{k}\right) }} \\
&=&\frac{W\left( 1,W\left( \psi _{\nu _{m+1}}^{\left( N_{m}/\nu _{k}\right)
},\psi _{\nu _{k}}^{\left( N_{m}/\nu _{k}\right) }\right) \mid x\right) }{%
\psi _{\nu _{k}}^{\left( N_{m}/\nu _{k}\right) }(x)W\left( \psi _{\nu
_{m+1}}^{\left( N_{m}/\nu _{k}\right) },\psi _{\nu _{k}}^{\left( N_{m}/\nu
_{k}\right) }\mid x\right) }  \notag \\
&=&-\frac{W^{\prime }\left( \psi _{\nu _{m+1}}^{\left( N_{m}/\nu _{k}\right)
},\psi _{\nu _{k}}^{\left( N_{m}/\nu _{k}\right) }\mid x\right) }{\psi _{\nu
_{k}}^{\left( N_{m}/\nu _{k}\right) }(x)W\left( \psi _{\nu _{m+1}}^{\left(
N_{m}/\nu _{k}\right) },\psi _{\nu _{k}}^{\left( N_{m}/\nu _{k}\right) }\mid
x\right) },  \notag
\end{eqnarray}%
that is, using the Wronskian theorem \cite{messiah}

\begin{equation}
\psi _{\nu _{k}}^{\left( N_{m+1}\right) }(x)\sim \frac{\psi _{\nu
_{m+1}}^{\left( N_{m}/\nu _{k}\right) }(x)}{W\left( \psi _{\nu _{k}}^{\left(
N_{m}/\nu _{k}\right) },\psi _{\nu _{m+1}}^{\left( N_{m}/\nu _{k}\right)
}\mid x\right) }\sim \frac{1}{\psi _{\nu _{k}}^{\left( N_{m}/\nu _{k}\right)
}}
\end{equation}

The property Eq(\ref{imageseed}) is also verified for $\left( m+1\right) $%
-step DBT, which ensures that it is verified for any number of steps.

\section{Degenerate chains}

Consider first the case of a two-step DBT with twice the use of the same
seed function. We have from Eq(\ref{psinunu})

\begin{equation}
V^{\left( \nu _{1},\nu _{1}\right) }(x)=V^{\left( \nu _{1}\right)
}(x)-2\left( \log \psi _{\nu _{1}}^{\left( \nu _{1}\right) }(x)\right)
^{\prime \prime }=V^{\left( \nu _{1}\right) }(x)+2\left( \log \psi _{\nu
_{1}}(x)\right) ^{\prime \prime }=V(x)  \label{Vnunu}
\end{equation}%
and the final potential coincides with the initial potential. The image of
the eigenfunction $\psi _{\lambda }$, $\lambda \neq \nu _{1}$, is given by

\begin{eqnarray}
\psi _{\lambda }^{\left( \nu _{1},\nu _{1}\right) }\left( x\right) &=&\frac{%
W\left( \psi _{\lambda }^{\left( \nu _{1}\right) },\psi _{\nu _{1}}^{\left(
\nu _{1}\right) }\mid x\right) }{\psi _{\nu _{1}}^{\left( \nu _{1}\right)
}\left( x\right) }\sim \frac{W\left( W\left( \psi _{\lambda },\psi _{\nu
_{1}}\right) /\psi _{\nu _{1}},1/\psi _{\nu _{1}}\mid x\right) }{1/\psi
_{\nu _{1}}\left( x\right) }  \label{psilambdanunu} \\
&\sim &\frac{W^{\prime }\left( \psi _{\lambda },\psi _{\nu _{1}}\mid
x\right) }{\psi _{\nu _{1}}\left( x\right) },  \notag
\end{eqnarray}%
that is, using the Wronskian theorem

\begin{equation}
\psi _{\lambda }^{\left( \nu _{1},\nu _{1}\right) }\left( x\right) \sim \psi
_{\lambda }\left( x\right) .  \label{psilambdanunu2}
\end{equation}

As for the image of $\psi _{\nu _{1}}$, it is

\begin{equation}
\psi _{\nu _{1}}^{\left( \nu _{1},\nu _{1}\right) }\left( x\right) \sim 
\frac{1}{\psi _{\nu _{1}}^{\left( \nu _{1}\right) }\left( x\right) }\sim
\psi _{\nu _{1}}\left( x\right) .  \label{psinununu}
\end{equation}

These results prove that, in the two-step case, using twice the same seed
function in a chain of DBT gives the identity transformation. As expected
intuitively, this result is generalizable to higher order chains. Namely, we
have

\begin{lem}\label{lem2}

By completing a chain by state-deleting DBT associated to some of
the previously used seed functions we obtain the images of a reduced chain
of $\left( m-k\right) $ steps where these seed functions
associated to $\nu _{i_{1}},...,\nu _{i_{k}}$ have been suppressed

\begin{equation}
\left\{ 
\begin{array}{c}
V^{\left( N_{m},\nu _{i_{1}},...,\nu _{i_{k}}\right) }\left( x\right)
=V^{\left( N_{m}/\nu _{i_{1}},...,\nu _{i_{k}}\right) }\left( x\right) \\ 
\psi _{\lambda }^{\left( N_{m},\nu _{i_{1}},...,\nu _{i_{k}}\right) }\left(
x\right) \sim \psi _{\lambda }^{\left( N_{m}/\nu _{i_{1}},...,\nu
_{i_{k}}\right) }\left( x\right) .%
\end{array}%
\right.  \label{lemma2}
\end{equation}
\end{lem}
\textit{Proof}:
Consider first a three step case $m=3$ where $N_{2}=\left( \nu _{1},\nu
_{2}\right) $ $N_{3}=\left( N_{2},\nu _{1}\right) $. We have from Eq(\ref%
{foDBTwronsk}) and Eq(\ref{potnstep})

\begin{equation}
V^{\left( N_{3}\right) }(x)V(x)-2\left( \log W\left( \psi _{\nu _{1}},\psi
_{\nu _{2}}\mid x\right) \psi _{\nu _{1}}^{\left( N_{2}\right) }\left(
x\right) \right) ^{\prime \prime }
\end{equation}

Using the fact that 

\begin{equation}
\psi _{\nu _{1}}^{\left( N_{2}\right) }\left( x\right) \sim \frac{\psi _{\nu
_{2}}\left( x\right) }{W\left( \psi _{\nu _{1}},\psi _{\nu _{2}}\mid
x\right) },
\end{equation}%
we obtain

\begin{equation}
V^{\left( N_{3}\right) }\left( x\right) =V^{\left( \nu _{2}\right) }\left(
x\right) .
\end{equation}

For $\lambda \neq \nu _{1},\nu _{2}$, we have, using Lemma \ref{lem1} and the
standard Crum formula

\begin{eqnarray}
\psi _{\lambda }^{\left( N_{3}\right) }\left( x\right) &=&\frac{W\left( \psi
_{\lambda }^{\left( N_{2}\right) },\psi _{\nu _{1}}^{\left( N_{2}\right)
}\mid x\right) }{\psi _{\nu _{1}}^{\left( N_{2}\right) }\left( x\right) }=%
\frac{W\left( \psi _{\lambda }^{\left( \nu _{2},\nu _{1}\right) },\psi _{\nu
_{1}}^{\left( \nu _{2},\nu _{1}\right) }\mid x\right) }{\psi _{\nu
_{1}}^{\left( \nu _{2},\nu _{1}\right) }\left( x\right) } \\
&\sim &\frac{W\left( W\left( \psi _{\lambda }^{\left( \nu _{2}\right) },\psi
_{\nu _{1}}^{\left( \nu _{2}\right) }\right) /\psi _{\nu _{1}}^{\left( \nu
_{2}\right) },1/\psi _{\nu _{1}}^{\left( \nu _{2}\right) }\mid x\right) }{%
1/\psi _{\nu _{1}}^{\left( \nu _{2}\right) }\left( x\right) }\sim \frac{%
W^{\prime }\left( \psi _{\lambda }^{\left( \nu _{2}\right) },\psi _{\nu
_{1}}^{\left( \nu _{2}\right) }\mid x\right) }{\psi _{\nu _{1}}^{\left( \nu
_{2}\right) }\left( x\right) }.  \notag
\end{eqnarray}

With the Wronskian theorem, we then obtain

\begin{equation}
\psi _{\lambda }^{\left( \nu _{1},\nu _{2},\nu _{1}\right) }\left( x\right)
\sim \psi _{\lambda }^{\left( \nu _{2}\right) }\left( x\right) .
\end{equation}

We can also note that, in a coherent way, using Eq(\ref{psilambdanunu2}) we
can write

\begin{equation}
\psi _{\lambda }^{\left( \nu _{1},\nu _{1},\nu _{2}\right) }\left( x\right) =%
\frac{W\left( \psi _{\lambda }^{\left( \nu _{1},\nu _{1}\right) },\psi _{\nu
_{2}}^{\left( \nu _{1},\nu _{1}\right) }\mid x\right) }{\psi _{\nu
_{2}}^{\left( \nu _{1},\nu _{1}\right) }\left( x\right) }=\frac{W\left( \psi
_{\lambda },\psi _{\nu _{2}}\mid x\right) }{\psi _{\nu _{2}}\left( x\right) }%
=\psi _{\lambda }^{\left( \nu _{2}\right) }\left( x\right)
\end{equation}%
and with Lemma \ref{lem1}

\begin{equation}
\psi _{\lambda }^{\left( \nu _{2},\nu _{1},\nu _{1}\right) }\left( x\right) =%
\frac{W\left( \psi _{\lambda }^{\left( \nu _{2},\nu _{1}\right) },\psi _{\nu
_{1}}^{\left( \nu _{2},\nu _{1}\right) }\mid x\right) }{\psi _{\nu
_{1}}^{\left( \nu _{2},\nu _{1}\right) }\left( x\right) }=\frac{W\left( \psi
_{\lambda }^{\left( \nu _{1},\nu _{2}\right) },\psi _{\nu _{1}}^{\left( \nu
_{1},\nu _{2}\right) }\mid x\right) }{\psi _{\nu _{1}}^{\left( \nu _{1},\nu
_{2}\right) }\left( x\right) }=\psi _{\lambda }^{\left( N_{3}\right) }\left(
x\right) =\psi _{\lambda }^{\left( \nu _{2}\right) }\left( x\right) .
\end{equation}

It remains to consider the cases $\lambda =\nu _{1}$ and $\lambda =\nu _{2}$%
. In the first case, we have

\begin{equation}
\psi _{\nu _{1}}^{\left( N_{3}\right) }\left( x\right) \sim \frac{1}{\psi
_{\nu _{1}}^{\left( N_{2}\right) }\left( x\right) }\sim \psi _{\nu
_{1}}^{\left( \nu _{2}\right) }(x)=\frac{\psi _{\nu _{2}}\left( x\right) }{%
W\left( \psi _{\nu _{1}},\psi _{\nu _{2}}\mid x\right) }
\end{equation}%
and in the second case

\begin{eqnarray}
\psi _{\nu _{2}}^{\left( N_{3}\right) }\left( x\right) &=&\frac{W\left( \psi
_{\nu _{2}}^{\left( N_{2}\right) },\psi _{\nu _{1}}^{\left( N_{2}\right)
}\mid x\right) }{\psi _{\nu _{1}}^{\left( N_{2}\right) }\left( x\right) }%
\sim \frac{W\left( 1/\psi _{\nu _{2}}^{\left( \nu _{1}\right) },1/\psi _{\nu
_{1}}^{\left( \nu _{2}\right) }\mid x\right) }{1/\psi _{\nu _{1}}^{\left(
\nu _{2}\right) }\left( x\right) } \\
&\sim &\frac{W\left( \frac{\psi _{\nu _{1}}}{W\left( \psi _{\nu _{2}},\psi
_{\nu _{1}}\right) },\frac{\psi _{\nu _{2}}}{W\left( \psi _{\nu _{1}},\psi
_{\nu _{2}}\right) }\mid x\right) }{\frac{\psi _{\nu _{2}}\left( x\right) }{%
W\left( \psi _{\nu _{1}},\psi _{\nu _{2}}\mid x\right) }},  \notag
\end{eqnarray}%
that is,

\begin{equation}
\psi _{\nu _{2}}^{\left( N_{3}\right) }\left( x\right) \sim \frac{1}{\psi
_{\nu _{2}}\left( x\right) }\sim \psi _{\nu _{2}}^{\left( \nu _{2}\right)
}\left( x\right) .
\end{equation}

The lemma is then verified in the $3$-step case.

Consider now the general case $N_{m}=\left( \nu _{1},...,\nu _{m}\right) $
and apply to $V^{\left( N_{m}\right) }$ the DBT $A(w_{\nu _{i_{1}}}^{\left(
N_{m}\right) })$. The resulting potential is

\begin{eqnarray}
\qquad V^{\left( N_{m},\nu _{i_{1}}\right) }\left( x\right) &=&V^{\left(
N_{m}\right) }\left( x\right) -2\left( \log \left( \psi _{\nu
_{i_{1}}}^{\left( N_{m}\right) }\left( x\right) \right) \right) ^{\prime
\prime }=V-2\left( \log \left( W^{\left( N_{m}\right) }\left( x\right) \psi
_{\nu _{i_{1}}}^{\left( N_{m}\right) }\left( x\right) \right) \right)
^{\prime \prime } \\
&=&V\left( x\right) -2\left( \log \left( W^{\left( N_{m}/\nu _{i_{1}}\right)
}\left( x\right) \right) \right) ^{\prime \prime }=V^{\left( N_{m}/\nu
_{i_{1}}\right) }\left( x\right) .  \notag
\end{eqnarray}

The recurrence is immediate and if we apply a chain of $k$ DBT based on the
seed functions $\psi _{\nu _{i_{1}}}^{\left( N_{m}\right) },...,$ $\psi
_{\nu _{i_{k}}}^{\left( N_{m}\right) }$ of $V^{\left( N_{m}\right) },$ $%
1\leq i_{1}<...<i_{k}\leq m$, the final extended potential can be written as

\begin{equation}
V^{\left( N_{m},\nu _{i_{1}},...,\nu _{i_{k}}\right) }\left( x\right)
=V^{\left( N_{m}/\nu _{i_{1}},...,\nu _{i_{k}}\right) }\left( x\right) .
\end{equation}

We know that $\psi _{\lambda }^{\left( N_{m}/\nu _{i_{1}},...,\nu
_{i_{k}}\right) }\left( x\right) $ is an eigenfunction of $V^{\left(
N_{m},\nu _{i_{1}},...,\nu _{i_{k}}\right) }\left( x\right) $ for the
eigenvalue $E_{\lambda }$ but we  still have to verify that it can be
identified with $\psi _{\lambda }^{\left( N_{m},\nu _{i_{1}},...,\nu
_{i_{k}}\right) }\left( x\right) $.

Consider first the case $\lambda =\nu _{j}\in N_{m}$. The image of $\psi
_{\nu _{j}}^{\left( N_{m}\right) }$, where $1\leq j\leq m$, by the DBT $%
A\left( \psi _{\nu _{i_{1}}}^{\left( N_{m}\right) }\right) ,$ $j\neq i_{1},$
is

\begin{equation}
\psi _{\nu _{j}}^{\left( N_{m},\nu _{i_{1}}\right) }\left( x\right) =\frac{%
W\left( \psi _{\nu _{j}}^{\left( N_{m}\right) },\psi _{\nu _{i_{1}}}^{\left(
N_{m}\right) }\mid x\right) }{\psi _{\nu _{i_{1}}}^{\left( N_{m}\right)
}\left( x\right) }.
\end{equation}

Using Lemma \ref{lem1}, it can be rewritten as

\begin{equation}
\psi _{\nu _{j}}^{\left( N_{m},\nu _{i_{1}}\right) }\left( x\right) \sim 
\frac{W\left( W^{\left( N_{m}/\nu _{j}\right) },W^{\left( N_{m}/\nu
_{i_{1}}\right) }\mid x\right) }{W^{\left( N_{m}\right) }\left( x\right)
W^{\left( N_{m}/\nu _{i_{1}}\right) }\left( x\right) }.
\end{equation}%
and using the Jacobi identity for Wronskians \cite{muir}

\begin{equation}
W^{\left( N_{m}\right) }\left( x\right) =\frac{W\left( W^{\left( N_{m}/\nu
_{j}\right) },W^{\left( N_{m}/\nu _{i_{1}}\right) }\mid x\right) }{W^{\left(
N_{m}/\nu _{j},\nu _{i_{1}}\right) }\left( x\right) },
\end{equation}%
we obtain

\begin{equation}
\psi _{\nu _{j}}^{\left( N_{m},\nu _{i_{1}}\right) }\left( x\right) \sim 
\frac{W^{\left( N_{m}/\nu _{j},\nu _{i_{1}}\right) }\left( x\right) }{%
W^{\left( N_{m}/\nu _{i_{1}}\right) }\left( x\right) }\sim \psi _{\nu
_{j}}^{\left( N_{m}/\nu _{i_{1}}\right) }\left( x\right) .
\end{equation}

The generalization is immediate and we have for $j\neq i_{1},...,i_{k}$%
\begin{equation}
\psi _{\nu _{j}}^{\left( N_{m},\nu _{i_{1}},...,\nu _{i_{k}}\right) }\left(
x\right) \sim \frac{W^{\left( N_{m}/\nu _{j},\nu _{i_{1}},...,\nu
_{i_{k}}\right) }\left( x\right) }{W^{\left( N_{m}/\nu _{i_{1}},...,\nu
_{i_{k}}\right) }\left( x\right) }\sim \psi _{\nu _{j}}^{\left( N_{m}/\nu
_{i_{1}},...,\nu _{i_{k}}\right) }\left( x\right) .
\end{equation}

In the case where $\lambda \notin N_{m}$, using Lemma \ref{lem1} and applying a DBT $%
A\left( \psi _{\nu _{i_{1}}}^{\left( N_{m}\right) }\right) $ to $\psi
_{\lambda }^{\left( N_{m}\right) }\left( x\right) ,$ we deduce

\begin{equation}
\psi _{\lambda }^{\left( N_{m},\nu _{i_{1}}\right) }\left( x\right) =\frac{%
W\left( \psi _{\lambda }^{\left( N_{m}\right) },\psi _{\nu _{i_{1}}}^{\left(
N_{m}\right) }\mid x\right) }{\psi _{\nu _{i_{1}}}^{\left( N_{m}\right)
}\left( x\right) }\sim \frac{W\left( W\left( \psi _{\lambda }^{\left(
N_{m}/\nu _{i_{1}}\right) },\psi _{\nu _{i_{1}}}^{\left( N_{m}/\nu
_{i_{1}}\right) }\right) /\psi _{\nu _{i_{1}}}^{\left( N_{m}/\nu
_{i_{1}}\right) },1/\psi _{\nu _{i_{1}}}^{\left( N_{m}/\nu _{i_{1}}\right)
}\mid x\right) }{1/\psi _{\nu _{i_{1}}}^{\left( N_{m}/\nu _{i_{1}}\right)
}\left( x\right) },
\end{equation}%
that is, using the Wronskian theorem%
\begin{equation}
\psi _{\lambda }^{\left( N_{m},\nu _{i_{1}}\right) }\left( x\right) \sim 
\frac{W^{\prime }\left( \psi _{\lambda }^{\left( N_{m}/\nu _{i_{1}}\right)
},\psi _{\nu _{i_{1}}}^{\left( N_{m}/\nu _{i_{1}}\right) }\mid x\right) }{%
\psi _{\nu _{i_{1}}}^{\left( N_{m}/\nu _{i_{1}}\right) }\left( x\right) }%
\sim \psi _{\lambda }^{\left( N_{m}/\nu _{i_{1}}\right) }\left( x\right) .
\end{equation}

Here again the generalization is immediate and we can write

\begin{equation}
\psi _{\lambda }^{\left( N_{m},\nu _{i_{1}},...,\nu _{i_{k}}\right) }\left(
x\right) \sim \psi _{\lambda }^{\left( N_{m}/\nu _{i_{1}},...,\nu
_{i_{k}}\right) }\left( x\right) ,
\end{equation}%
which concludes the proof of Lemma \ref{lem2}.

\section{Disconjugacy and regular extensions}

From the original potential $V$, the DBT $A\left( w_{\nu }\right) $ generates a new potential $%
V^{\left( \nu \right) }$ formally isospectral to the original one and its
eigenfunctions are directly obtained from those of $V$ via Eq(\ref{foDBT}).
Nevertheless, in general, $w_{\nu }(x)$ and then the transformed potential $%
V^{\left( \nu \right) }(x)$ are singular at the nodes of $\psi _{\nu }(x)$.
For instance, if $\psi _{n}(x)$ ($\nu =n$) is a bound state of $\widehat{H}$%
, the Sturm oscillation theorem \cite{hartman,bocher} implies that $%
V^{\left( n\right) }$ is regular only when $n=0$, that is, when the seed
function is the ground state of $\widehat{H}$. This corresponds exactly to
the usual SUSY partnership in quantum mechanics \cite{cooper,Dutt}.

We can however envisage to use as seed function any other regular solution
of Eq(\ref{edr4}) as long as it has no zeros on the considered real interval $%
I$ and transformed eigenfunction satisfies the required boundary
conditions, even if this seed function does not correspond to a physical
state.

To control the regularity of $w_{\nu }$ we can make use of the disconjugacy
properties of the Schr\"{o}dinger equation for negative eigenvalues. In the
SUSY QM frame, this possibility has been first envisaged by Sukumar \cite%
{sukumar}.

A second order differential equation like Eq(\ref{EdS}) is said to be 
\textit{disconjugated} on $I\subset \mathbb{R}$ ($V(x)$ is supposed to be
continuous on $I$) if every solution of this equation has at most one zero
on $I$ \cite{hartman,coppel,derr}. As it is well known, this zero is
necessarily simple and at this value the considered solution changes its
sign. For a closed or open interval $I$, the disconjugacy of Eq(\ref{EdS})
is equivalent to the existence of solutions of this equation which are
everywhere non zero on $I$ \cite{hartman,coppel,derr}.

We will make use of the following result \cite{hartman,coppel}:

\ 

\begin{thm}[Disconjugacy theorem] If there exists
a continuously differentiable solution on $I$ of the Riccati
inequation

\begin{equation}
-w^{\prime }(x)+w^{2}(x)\leq G(x),  \label{Ricineq}
\end{equation}%
then the equation

\begin{equation}
\psi ^{\prime \prime }(x)-G(x)\psi (x)=0  \label{EdSb}
\end{equation}%
is disconjugated on $I$.
\end{thm}

 In our case Eq(\ref{EdSb}) is the Schr\"{o}dinger equation Eq(\ref{EdS})
with $G(x)=V(x)-E_{\lambda }$ and if $E_{\lambda }\leq E_{0}=0$, we have

\begin{equation}
-w_{0}^{\prime }(x)+w_{0}^{2}(x)\leq V(x)-E_{\lambda },  \label{inegdeconj}
\end{equation}%
$w_{0}(x)$ being continuously differentiable on $I$. The above theorem
ensures the existence of nodeless solutions of Eq(\ref{EdS}) for the
eigenvalue $E_{\lambda }$. To prove that a given solution $\phi (x)$ belongs
to this category, it is sufficient to determine the signs of the boundaries
of $I$. If they are identical then $\phi $ is nodeless and if they are
opposite, then $\phi $ presents a unique zero on $I$. In the first case $%
V(x)+2v^{\prime }(x)$, where $v(x)=-\phi ^{\prime }(x)/\phi (x)$,
constitutes a regular (quasi)isospectral extension of $V(x)$.\ The \textit{%
disconjugacy sector} of Eq(\ref{EdS}) corresponds then to the values of the
spectral parameter $E_{\lambda }\leq E_{0}$.

\section{Regular chains of DBT and Krein-Adler theorem}

The question of the link between the regularity of the successive extensions
obtained by a chain of DBT and the choice of the associated seed functions
is natural. In his seminal paper, Crum \cite{crum,matveev} has first
established that if we take a sequence $N_{m}=\left( 0,...,m-1\right) =M$
corresponding to seed functions which are consecutive eigenstates starting
from the fundamental one and that we call a \textit{complete chain of DBT},
then 
\begin{equation}
V^{\left( M\right) }(x)=V(x)+2\left( \log W\left( \psi _{0},...,\psi
_{m-1}\mid x\right) \right) ^{\prime \prime }
\end{equation}%
is regular and its spectrum is given by

\begin{equation}
\left\{ 
\begin{array}{c}
E_{k}^{\left( M\right) }=E_{k+m} \\ 
\phi _{k}^{\left( M\right) }(x)=\psi _{k+m}^{\left( M\right) }(x)=\frac{%
W\left( \psi _{0},...,\psi _{m-1},\psi _{k}\mid x\right) }{W\left( \psi
_{0},...,\psi _{m-1}\mid x\right) }%
\end{array}%
\right. ,\quad k\geq 0.
\end{equation}

This means that although each of the $\psi _{i},$ $1\leq i\leq m-1,$ has
exactly $i$ nodes, the Wronskian of the family $\left( \psi _{0},...,\psi
_{m-1}\right) $ is, as for it, free of zero.

It can be very easily understood by noting that the action on the spectrum
of the successive DBT based on such eigenstates corresponds to suppress at
each step the level associated to the used seed function and the DBT are
said \textit{state-deleting}. In this perspective, since $\psi _{j}^{\left(
N_{j}\right) }(x),$ $N_{j}=\left( 0,...,j-1\right) ,$ is the fundamental
eigenstate of $V^{\left( N_{j}\right) }(x)$, the chain can be viewed as a
succession of SUSY QM partnerships and all the extensions along the chain
are regular. This is exactly the argument at the basis of the construction
of hamiltonians hierarchy in SUSY QM \cite{cooper,Dutt}.

Krein \cite{krein} and later Adler \cite{adler} have enlarged this
regularity property to lacunary chains of state-deleting DBT.

\begin{thm}[Krein-Adler]

Consider a chain of DBT characterized by the $m$-uple $N_{m}=\left( n_{1},,...,n_{m}\right) $ of positive integers such
that the corresponding seed functions $\psi _{n_{i}}$ are
eigenstates of the initial potential. 

The extended potential $V^{\left( N_{m}\right) }(x)$ is
regular if and only if the sequence $N_{m}$ is constituted by fragments of an
even number of consecutive positive integers, except the first one which is
of arbitrary length. In other words, the spectrum of the final extension $%
V^{\left( N_{m}\right) }(x)$ must contain only \textit{even gaps}
(gaps constituted by an even number of consecutive missing levels). This
condition is equivalently described by the Krein condition

\begin{equation}
(n-n_{1})(n-n_{2})...(n-n_{m})\geq 0,\ \forall n\in N.
\end{equation}
\end{thm}
Such a lacunary chain is said of the \textit{Krein-Adler type}.
In \cite{samsonov2}, Samsonov has still extended the Krein-Adler result,
showing the possibility to employ sets of two juxtaposed eigenfunctions
whose associated eigenvalues are taken between two consecutive energy
levels. These results have been used extensively in the context of higher
order SUSY (see for instance \cite%
{sukumar,sukumar2,andrianov,andrianov2,bagrov,bagrov1,samsonov,samsonov1,samsonov2,fernandez,fernandez2,fernandez3,mielnik}%
). .

Recently we have proven that for the specific class of translationally shape
invariant potentials of the second category, it is possible to build chains
of regular DBT of arbitrary length by using eigenfunctions in the
disconjugacy sector \cite{gomez5,grandati5,quesne5} associated to
exceptional Laguerre or Jacobi polynomials.

\section{Translationally Shape Invariant Potentials (TSIP)}

Consider a potential $V(x;\alpha )$ which depends upon a (multi)parameter $%
\alpha \in \mathbb{R}^{N}$ and with a (finite or infinite) bound state
spectrum $\left( E_{n},\psi _{n}\right) _{n\geq 0}$, the ground level being
supposed to be zero: $E_{0}(\alpha )=0$. In the framework of SUSY QM, such a
potential\ is said to be \textit{shape invariant (SIP)} \cite%
{cooper,Dutt,gendenshtein} if its SUSY partner

\begin{equation}
V^{\left( 0\right) }(x;\alpha )=V(x;\alpha )+2w_{0}^{\prime }(x;\alpha ),
\end{equation}%
keeps the same functional form as the initial potential. Namely

\begin{equation}
V^{\left( 0\right) }(x;\alpha )=V(x;f(\alpha ))+R(\alpha ),  \label{SI}
\end{equation}%
$R\left( \alpha \right) \in \mathbb{R}$ and $f(\alpha )\in \mathbb{R}^{N}$
being two given functions of $\alpha $.

In this case, it can be shown\ \cite{cooper,Dutt,gendenshtein} that the
complete bound state energy spectrum of $\widehat{H}(\alpha )=-\frac{d^{2}}{%
dx^{2}}+V(x;\alpha )$ is given by:

\begin{equation}
E_{n}(\alpha )=\sum_{k=0}^{n-1}R(\alpha _{k}),  \label{spectreSIP}
\end{equation}%
where $\alpha _{k}=f^{\left( k\right) }(\alpha )=\overset{\text{k times}}{%
\overbrace{f\circ ...\circ f}}(\alpha )$.

As for the corresponding eigenstates, they can be written as

\begin{equation}
\psi _{n}(x;\alpha )\sim \widehat{A}^{+}(\alpha )\psi _{n-1}(x;\alpha
_{1})\sim \widehat{A}^{+}(\alpha )...\widehat{A}^{+}(\alpha _{n-1})\psi
_{0}(x;\alpha _{n}),  \label{spectreSIP2}
\end{equation}%
where $\widehat{A}^{+}(\alpha )=-\frac{d}{dx}+w_{0}(x;\alpha )$. For the
corresponding RS functions the above differential relation becomes an
algebraic one

\begin{equation}
w_{n}(x;\alpha )=w_{0}(x;\alpha )-\frac{E_{n}(\alpha )}{w_{0}(x;\alpha
)+w_{n-1}(x;\alpha _{1})},  \label{eqfond3}
\end{equation}%
which allows to write the excited states as terminating continued fractions
in terms of $w_{0}$ \cite{grandati}.

When $f$ is a simple translation $f(\alpha )=\alpha +\varepsilon ,\
\varepsilon =\left( \varepsilon ^{\left( 1\right) },...,\varepsilon ^{\left(
N\right) }\right) \in \mathbb{R}^{N}$, $V$ is said to be \textit{%
translationally shape invariant} and we call it a \textit{TSIP}. Without
loss of generality, we can take $\varepsilon ^{\left( k\right) }=\pm 1$ or $%
0 $. For all the known TSIP we have $\alpha \in 
\mathbb{R}
$ (first category TSIP) or $\alpha =\left( \alpha ^{\left( 1\right) },\alpha
^{\left( 2\right) }\right) \in 
\mathbb{R}
^{2}$ (second category TSIP) \cite{cooper,Dutt,grandati}.

Since $R\left( \alpha \right) =E_{1}(\alpha )$, Eq(\ref{spectreSIP}) can be
rewritten as

\begin{equation}
E_{n}(\alpha )=\sum_{i=0}^{n-1}E_{1}(\alpha _{i})
\end{equation}%
and the dispersion relation, that is, $E_{n}(\alpha )$ as a function of $n$
and $\alpha $, satisfies then the characteristic identity

\begin{equation}
E_{n+m}(\alpha )-E_{m}(\alpha )=E_{n}(\alpha _{m}),  \label{eqcaracenerTSIP}
\end{equation}%
which can be used to extend the dispersion relation to every integer values
of $n$. We easily verify that this identity is satisfied as soon as $%
E_{n}(\alpha )$ is of the form

\begin{equation}
E_{n}(\alpha )=\pm \left( g(\alpha )-g(\alpha _{n})\right) ,
\end{equation}%
$g(\alpha )$ being an arbitrary function of $\alpha $.

The set of TSIP contains all the potentials classically known to be exactly
solvable, i.e. for which we know explicitly the dispersion relation and whose
the eigenfunctions can be expressed in closed analytical form in terms of
elementary transcendental functions, namely the harmonic, isotonic, Morse,
Kepler-Coulomb, Eckart, Darboux-P\"{o}schl-Teller hyperbolic and
trigonometric and Rosen-Morse hyperbolic and trigonometric potentials. Until
recently it was commonly believed \ that these potentials were the only ones
to possess the translational shape invariance property. However, it has been
shown recently that they are in fact only \textit{primary TSIP (PTSIP)} from
which it is possible in some cases to build infinite towers of secondary
TSIP (STSIP) which are extensions of the previous ones and which share the
same translational shape invariance properties \cite{grandati5}.

An important feature of the the PTSIP is that their eigenfunctions $\psi
_{n} $ are equal, up to a gauge factor, to classical orthogonal polynomials
in an appropriate variable $z$ (which can be $n$ dependent) and we call $%
\psi _{n}$ a \textit{quasi-polynomial} in this variable. The associated RS
functions are rational functions of $z$. As for the function
defined above, it is given by $g(\alpha )=\alpha ^{2}$ or $g(\alpha
)=1/\alpha ^{2}$, \cite{grandati}.

Note that in general, the range of values of the multiparameter $\alpha $ is
subject to limiting constraints of the type $\alpha \in U\subset 
\mathbb{R}
^{N}$, that is, $\alpha ^{\left( i\right) }\in U^{\left( i\right) },\quad
U^{\left( i\right) }\subset 
\mathbb{R}
$, in order to have a appropriate regular behaviour for $V(x;\alpha )$ and
to ensure that the $\psi _{n}(x;\alpha )$ satisfy the required Dirichlet
boundary conditions.

\section{ Discrete symmetries, rational extensions of TSIP and state-adding
DBT}

The primary TSIP potentials possess specific discrete symmetries $\Gamma
_{i} $ which act in the space of parameters and which are covariance
transformations for the considered potential

\begin{equation}
\left\{ 
\begin{array}{c}
\alpha \overset{\Gamma _{i}}{\rightarrow }\alpha _{i} \\ 
V(x;\alpha )\overset{\Gamma _{i}}{\rightarrow }V(x;\alpha _{i})=V(x;\alpha
)+\delta _{i}\left( \alpha \right) .%
\end{array}%
\right.  \label{sym}
\end{equation}

In the space of parameters (which is two-dimensional for the second category
and one-dimensional for the first), the $\Gamma _{i}$ are reflections with
respect to the coordinate axis or combinations of them. For the second
category potentials (for which the parameter is two-dimensional) we then
have three different symmetries $\Gamma _{+}$, $\Gamma _{-}$ and $\Gamma
_{3}=\Gamma _{+}\circ $ $\Gamma _{-}$, while for the first category ones
(for which the parameter is one-dimensional) we have only one symmetry $%
\Gamma _{3}$.

$\Gamma _{i}$ transforms an excited eigenstate $\psi _{n}$ into a unphysical
RS function $\phi _{n,i}(x;\alpha )=\Gamma _{i}\left( \psi _{n}(x;\alpha
)\right) $ in the disconjugacy sector of $V(x;\alpha )$, that is, associated
to the negative eigenvalue $\mathcal{E}_{n,i}(\alpha )=\Gamma _{i}\left(
E_{n}(\alpha )\right) =E_{n}(\alpha _{i})-\delta _{i}\left( \alpha \right)
<0.$

The $\Gamma _{i}$ symmetries preserve the functional structure of the
initial eigenstates and the $\phi _{n,i}$ are then also quasi-polynomials
with rational associated RS functions $v_{n,i}(x;\alpha )$.

If the transformed RS function $v_{n,i}$ is regular on $I$, it can be used
to build a regular extended potential (see Eq(\ref{pottrans}) and Eq(\ref%
{foDBT}))

\begin{equation}
V^{\left( n,i\right) }(x;\alpha )=V(x;\alpha )+2v_{n,i}^{\prime }(x;\alpha )
\end{equation}%
(quasi)isospectral to $V(x;\alpha )$ that we call a \textit{rational
extension} of $V$.

To the $\psi _{k\geq 0}$ correspond eigenstates of $V^{\left( n,i\right) }$
which are given by (see Eq(\ref{transfoback2}))

\begin{equation}
\left\{ 
\begin{array}{c}
w_{k}^{\left( n,i\right) }(x;\alpha )=-v_{n,i}(x;\alpha )+\left(
E_{k}(\alpha )-\mathcal{E}_{n,i}(\alpha )\right) /\left( v_{n,i}(x;\alpha
)-w_{k}(x;\alpha )\right) \\ 
\psi _{k}^{\left( n,i\right) }(x;\alpha )=\exp \left( -\int
dxw_{k}^{(n,i)}(x;\alpha )\right) \sim W\left( \phi _{n,i},\psi _{k}\mid
x\right) /\phi _{n,i}\left( x;\alpha \right) ,%
\end{array}%
\right.  \label{foext}
\end{equation}%
for the respective\ energies $E_{k}(\alpha )$.\ 

The nature of the isospectrality depends if $1/\phi _{n,i}(x;\alpha )$\
satisfies or not the appropriate Dirichlet boundary conditions. If it is the
case, then $1/\phi _{n,i}(x;\alpha )$ is a physical eigenstate of $\widehat{H%
}^{\left( n,i\right) }(\alpha )=-d^{2}/dx^{2}+V^{\left( n,i\right)
}(x;\alpha )$ for the eigenvalue $\mathcal{E}_{n,i}(\alpha )$ and we only
have quasi-isospectrality (or essential isospectrality) between $V(x;\alpha
) $\ and\ $V^{\left( n,i\right) }(x;\alpha )$, the spectrum of $V^{\left(
n,i\right) }$ containing a supplementary level $\mathcal{E}_{n,i}(\alpha )$
below $E_{0}=0$. The DBT $A\left( v_{n,i}\right) $ is then called a \textit{%
state-adding DBT}. If it is not the case, the isospectrality between $%
V^{\left( n,i\right) }(x;\alpha )$ and $V(x;\alpha )$ is strict.

It has been proven \cite{grandati3} that the regular $V^{\left( n,+\right) }$%
\ and\ $V^{\left( n,-\right) }$ are isospectral to the initial potential
while the regular $V^{\left( n,3\right) }$ are only quasi-isospectral to $V$%
. Moreover, the regular extensions $V^{\left( n,\pm \right) }$ inherit of
the shape invariance properties of the initial potential while the regular $%
V^{\left( n,3\right) }$ not. The DBT $A\left( v_{n,3}\right) $ is in fact a
reciprocal SUSY\ QM\ partnership

\begin{equation}
A\left( v_{n,3}\right) =A^{-1}\left( w_{0}^{\left( n\right) }\right) .
\label{resSUSY}
\end{equation}

In th following we consider specifically the extensions associated to the $%
\Gamma _{3}$ symmetry. It acts as

\begin{equation}
\left\{ 
\begin{array}{c}
\Gamma _{3}\left( V\left( x;\alpha \right) \right) =V\left( x;-\alpha
\right) =V\left( x;\alpha \right) +\delta (\alpha ) \\ 
\Gamma _{3}\left( \psi _{n}\left( x;\alpha \right) \right) =\psi _{n}\left(
x;-\alpha \right) =\phi _{n,3}(x;\alpha )%
\end{array}%
\right.  \label{gamma3}
\end{equation}%
(if $\left( \alpha ^{\left( 1\right) },\alpha ^{\left( 2\right) }\right)
=\left( \alpha ,\beta \right) $ then $-\alpha =\left( -\alpha ^{\left(
1\right) },-\alpha ^{\left( 2\right) }\right) $ and we note $\delta (\alpha
) $ for $\delta _{3}(\alpha )$).

For all the PTSIP we have

\begin{equation}
\mathcal{E}_{n,3}(\alpha )=E_{-(n+1)}(\alpha )<0  \label{E3}
\end{equation}%
and $\phi _{n,3}(x;\alpha )$ diverges at both limits of the definition
domain (which implies that $1/\phi _{n,3}$\ satisfies the required Dirichlet
boundary conditions).

This leads to fix

\begin{equation}
\psi _{-\left( n+1\right) }\left( x;\alpha \right) =\phi _{n,3}(x;\alpha ).
\label{phi3}
\end{equation}

This implies also the identity ($E_{0}(\alpha )=0$)

\begin{equation}
\delta (a)=E_{-(n+1)}(\alpha )-E_{n}(\alpha )=E_{-1}(\alpha ),  \label{delta}
\end{equation}%
with (see Eq(\ref{eqcaracenerTSIP}))

\begin{equation}
E_{-1}(\alpha )=-E_{1}(\alpha _{-1}).  \label{E-1}
\end{equation}

Since $\psi _{-\left( n+1\right) }$ is in the disconjugacy sector of $%
V\left( x;\alpha \right) $, the disconjugacy theorem implies that it is free
of nodes only if it has the same sign at the boundaries of the definition
interval, which is the case when $\alpha _{n}\in U$ and $n$ is odd.

\section{Generalized Krein-Adler theorem for TSIP}

\subsection{Complete chains of state-adding DBT for TSIP}

For the TSIP we can envisage to build complete chains of DBT based not only
on successive state-deleting DBT but also complete chains of DBT based on
state-adding DBT. These chains, characterized by $m$-uple of the type $%
-M=\left( -1,...,-m\right) $, correspond to several successive \textit{%
reversed SUSY partnerships} and are "mirror images" of the chains of
state-deleting DBT associated to the sequence $M=\left( 0,...,m-1\right) $.
The extended potentials thus generated satisfy also a reversed shape
invariance condition.

\begin{lem}\label{lem3}

Suppose that TSIP $V(x;\alpha )$ admits a discrete $\Gamma _{3}$ symmetry such that $\Gamma _{3}\left( \psi _{n}\left(
x;\alpha \right) \right) =\psi _{-\left( n+1\right) }\left( x;\alpha \right).$ If $\alpha _{-1}\in A$, the action of the DBT $A\left(
w_{-1}\right) $ on $V(x;\alpha )$  is a reverse SUSY
partnership and $V\left( x;\alpha \right) $ satisfies a reverse
shape invariance condition

\begin{equation}
V^{\left( -1\right) }\left( x;\alpha \right) =V\left( x;\alpha _{-1}\right)
+E_{-1}(\alpha ).  \label{revSUSYSI}
\end{equation}

When $A\left( w_{-1}\right) $ is state-adding, the
spectrum of the regular extended potential $V^{\left( -1\right) }\left(
x;\alpha \right) $ is given by

\begin{equation}
\left\{ 
\begin{array}{c}
E_{j}^{\left( -1\right) }(\alpha )=E_{j-1}(\alpha ) \\ 
\phi _{0}^{\left( -1\right) }\left( x;\alpha \right) =\psi _{-1}^{\left(
-1\right) }\left( x;\alpha \right) \sim 1/\psi _{-1}\left( x;\alpha \right)
\sim \psi _{0}\left( x;\alpha _{-1}\right) \\ 
\phi _{j+1}^{\left( -1\right) }\left( x;\alpha \right) =\psi _{j}^{\left(
-1\right) }\left( x;\alpha \right) \sim \frac{W\left( \psi _{j},\psi
_{-1}\mid x\right) }{\psi _{-1}\left( x;\alpha \right) }\sim \psi
_{j+1}\left( x;\alpha _{-1}\right)%
\end{array}%
,\quad j\geq 0.\right.  \label{revSUSYspec}
\end{equation}
\end{lem}

\textit{Proof:}
Acting on $V(x;\alpha )$ with the regular DBT $A\left( w_{-1}\right)
=A\left( v_{0,3}\right) $ and using Eq(\ref{gamma3}) and Eq(\ref{delta}), we
obtain%
\begin{eqnarray}
V^{\left( -1\right) }\left( x;\alpha \right) &=&V\left( x;\alpha \right)
+2w_{-1}^{\prime }\left( x;\alpha \right) =V\left( x;-\alpha \right)
-E_{-1}(\alpha )+2w_{0}^{\prime }\left( x;-\alpha \right)  \label{V-1} \\
&=&V^{\left( 0\right) }\left( x;-\alpha \right) -E_{-1}(\alpha ).  \notag
\end{eqnarray}

Applying the translational shape invariance property (see Eq(\ref{SI}) of $V$%
, we obtain

\begin{equation}
V^{\left( -1\right) }\left( x;\alpha \right) =V\left( x;\left( -\alpha
\right) _{1}\right) +E_{1}(-\alpha )-E_{-1}(\alpha )=V\left( x;-\alpha
_{-1}\right) +E_{1}(-\alpha )-E_{-1}(\alpha ),
\end{equation}%
or (see Eq(\ref{gamma3}))

\begin{equation}
V^{\left( -1\right) }\left( x;\alpha \right) =V\left( x;\alpha _{-1}\right)
-S(\alpha ),  \label{rsi}
\end{equation}%
with

\begin{equation}
S(\alpha )=E_{-1}(\alpha )-E_{-1}(\alpha _{-1})-E_{1}(-\alpha ).  \label{S}
\end{equation}

Eq(\ref{rsi}) ensures the regularity of $V^{\left( -1\right) }$ when\textit{%
\ }$\alpha _{-1}\in U$. It is a shape invariance condition for the reversed
SUSY partnership corresponding to the DBT $A\left( w_{-1}\right) $ based on
the "symmetrized ground state" $\psi _{-1}=\Gamma _{3}\left( \psi
_{0}\right) $.

\bigskip If $A\left( w_{-1}\right) $ is a state-adding DBT, the ground level
of $V^{\left( -1\right) }$ is $E_{-1}(\alpha )$ with a fundamental eigenstate

\begin{equation}
\phi _{0}^{\left( -1\right) }\left( x;\alpha \right) =\psi _{-1}^{\left(
-1\right) }\left( x;\alpha \right) =1/\psi _{-1}\left( x;\alpha \right) .
\label{psi-1-1}
\end{equation}

But the ground level of $V\left( x;\alpha _{-1}\right) -S(\alpha )$ is $%
E_{0}(\alpha _{-1})-S(\alpha )=-S(\alpha )$ with the corresponding
eigenstate $\psi _{0}\left( x;\alpha _{-1}\right) $ which (when\textit{\ }$%
\alpha _{-1}\in A$) is supposed to satisfy the required Dirichlet boundary
conditions. We then deduce

\begin{equation}
\left\{ 
\begin{array}{c}
S(\alpha )=-E_{-1}(\alpha )\geq 0 \\ 
\psi _{-1}^{\left( -1\right) }\left( x;\alpha \right) \sim \psi _{0}\left(
x;\alpha _{-1}\right) \sim 1/\psi _{-1}\left( x;\alpha \right) .%
\end{array}%
\right.
\end{equation}

As for the excited eigenstates $\phi _{j+1}^{\left( -1\right) }$ of this
extension, associated to the eigenvalues $E_{j\geq 0}(\alpha )$, they are
given by the usual Darboux-Crum formula Eq(\ref{foDBTwronsk})

\begin{equation}
\phi _{j+1}^{\left( -1\right) }\left( x;\alpha \right) =\psi _{j}^{\left(
-1\right) }\left( x;\alpha \right) \sim \frac{W\left( \psi _{j},\psi
_{-1}\mid x\right) }{\psi _{-1}\left( x;\alpha \right) }.
\end{equation}

Using Eq(\ref{rsi}) and Eq(\ref{eqcaracenerTSIP}), we see that $\psi
_{j}^{\left( -1\right) }\left( x;\alpha \right) $ is also an eigenstate of $%
\widehat{H}(\alpha _{-1})$ for the eigenvalue $E_{j+1}(\alpha )$, which
leads to

\begin{equation}
\phi _{j+1}^{\left( -1\right) }\left( x;\alpha \right) \sim \psi
_{j+1}\left( x;\alpha _{-1}\right) .
\end{equation}

The action of the reverse SUSY partnership $A(\psi _{-1})$ is completely
analogous to the usual SUSY partnership but with an opposite shift of the
parameter and the addition of a new ground level rather than the suppression
of the original one. We can then envisage to build a complete chain of
state-adding DBT, characterized by the $m$-uple $-M=\left( -1,...,-m\right) $, which corresponds to $m$ successive reversed SUSY partnerships.

\begin{lem}\label{lem4}

If $\alpha _{-j}\in U,\ \forall j\leq m$, the final
extension of a complete chain of $m$ state-adding DBT applied to
the TSIP is regular and satisfies the reverse shape invariance property

\begin{equation}
V^{\left( -M\right) }\left( x;\alpha \right) =V\left( x;\alpha _{-m}\right)
+E_{-m}(\alpha ).
\end{equation}
Its energy spectrum is given by $E_{j}^{\left( -M\right) }(\alpha
)=E_{j-m}(\alpha )$, $j\geq 0$, with the corresponding
eigenstates
\begin{equation}
\phi _{j}^{\left( -M\right) }(x;\alpha )=\psi _{j-m}^{\left( -M\right)
}(x;\alpha )\sim \psi _{j}\left( x;\alpha _{-m}\right) .
\end{equation}
\end{lem}

\textit{Proof.}
From Eq(\ref{revSUSYSI}), an immediate recurrence gives
\begin{equation}
V^{\left( -M\right) }\left( x;\alpha \right) =V\left( x;\alpha _{-m}\right)
+\sum_{i=0}^{-m+1}E_{-1}(\alpha _{i}).  \label{rsig}
\end{equation}
and Eq(\ref{eqcaracenerTSIP}) leads to
\begin{equation}
E_{-m}(\alpha )=E_{-m+1}(\alpha )+E_{-1}(\alpha
_{-m+1})=\sum_{i=0}^{-m+1}E_{-1}(\alpha _{i}),\quad m\geq 1,
\end{equation}%
we obtain thus
\begin{equation}
V^{\left( -M\right) }\left( x;\alpha \right) =V\left( x;\alpha _{-m}\right)
+E_{-m}(\alpha ).  \label{msteprevTSIP}
\end{equation}
Due to the quasi-isospectrality property of the state-adding DBT, the
Hamiltonian 
\begin{equation}
\widehat{H}^{\left( -M\right) }(\alpha )=-\frac{d^{2}}{dx^{2}}+V^{\left(
-M\right) }\left( x;\alpha \right) =\widehat{H}(\alpha _{-m})+E_{-m}(\alpha )
\label{hamext}
\end{equation}%
has the same energy spectrum as the initial potential with $m$ supplementary
levels at the values $E_{-m}(\alpha )<...<E_{-1}(\alpha )<0$. More precisely
the spectrum of $\widehat{H}^{\left( -M\right) }(\alpha )$ is given by
\begin{equation}
\left\{ 
\begin{array}{c}
E_{j}^{\left( -M\right) }(\alpha )=E_{-m+j}(\alpha ) \\ 
\phi _{j}^{\left( -M\right) }(x;\alpha )=\psi _{j-m}^{\left( -M\right)
}(x;\alpha )%
\end{array}%
\right. ,\quad j\geq 0.  \label{energies}
\end{equation}
The structural identity for the dispersion relation Eq(\ref{eqcaracenerTSIP}%
) gives
\begin{equation}
E_{k}^{\left( -M\right) }(\alpha )=E_{k}(\alpha _{-m})+E_{-m}(\alpha )
\label{eqcarac}
\end{equation}%
which combined with Eq(\ref{hamext}) and Eq(\ref{msteprevTSIP}), implies
\begin{equation}
\widehat{H}(\alpha _{-m})\phi _{k}^{\left( -M\right) }(x;\alpha
)=E_{k}(\alpha _{-m})\phi _{k}^{\left( -M\right) }(x;\alpha )
\end{equation}
This means in particular
\begin{equation}
\phi _{k}^{\left( -M\right) }(x;\alpha )=\psi _{k-m}^{\left( -M\right)
}(x;\alpha )\sim \psi _{k}(x;\alpha _{-m}).  \label{SIeigen}
\end{equation}%
when $\alpha _{-m}\in U$ in which case $\psi _{k}(x;\alpha _{-m})$ satisfies
the Dirichlet boundary conditions at the limit of the definition interval.

We observe that Eq(\ref{msteprevTSIP}) ensures the regularity of the potential $V^{\left(
-M\right) }$, which, applying the Crum formula Eq(\ref{potnstep2}), can also
be written as (see Eq(\ref{imageseed}))

\begin{equation}
V^{\left( -M\right) }\left( x;\alpha \right) =V\left( x;\alpha \right)
-2\left( \log W^{\left( -M\right) }\left( x;\alpha \right) \right) ^{\prime
\prime }.  \label{104}
\end{equation}

Consequently from Eq(\ref{msteprevTSIP}) and Eq(\ref{104}) we deduce the
nodeless character of the Wronskian $W^{\left( -M\right) }\left( x\right) $
for every value of $m$ for which $\alpha _{-m}\in U$.

Comparing Eq(\ref{msteprevTSIP}) and Eq(\ref{104}) gives also the following
identity for the eigenfunctions of the initial PTSIP

\begin{lem}\label{lem5}

If $\alpha _{-m}\in U$, then

\begin{equation}
\left\{ 
\begin{array}{c}
\psi _{k+m}(x;\alpha _{-m})\sim \frac{W^{\left( -M,k\right) }\left( x;\alpha
\right) }{W^{\left( -M\right) }\left( x;\alpha \right) }=\frac{W\left( \psi
_{-1},...,\psi _{-m},\psi _{k}\mid x;\alpha \right) }{W\left( \psi
_{-1},...,\psi _{-m}\mid x;\alpha \right) },\quad k\geq 0 \\ 
\\ 
\psi _{k+m}(x;\alpha _{-m})\sim \frac{W^{\left( -M/-k\right) }\left(
x;\alpha \right) }{W^{\left( -M\right) }\left( x;\alpha \right) }=\frac{%
W\left( \psi _{-1},...,\psi _{-k+1},\psi _{-k-1},...,\psi _{-m}\mid x;\alpha
\right) }{W\left( \psi _{-1},...,\psi _{-m}\mid x;\alpha \right) },\quad
0>k\geq -m.%
\end{array}%
\right.  \label{wronskid}
\end{equation}
\end{lem}

\textit{Proof.}
The eigenstates of $V^{\left( -M\right) }$ associated to the positive levels
($k\geq 0$) are given by (see Eq(\ref{etats n3}))
\begin{equation}
\psi _{k}^{\left( -M\right) }(x;\alpha )=\frac{W^{\left( -M,k\right) }\left(
x;\alpha \right) }{W^{\left( -M\right) }\left( x;\alpha \right) },\quad
k\geq 0,  \label{sawronsk1}
\end{equation}%
while the eigenstates of negative energies are given by (see Eq(\ref%
{imageseed}))\bigskip 
\begin{equation}
\psi _{-k}^{\left( -M\right) }(x;\alpha )=\frac{W^{\left( -M/-k\right)
}\left( x;\alpha \right) }{W^{\left( -M\right) }\left( x;\alpha \right) }%
,\quad k>0,  \label{sawronsk2}
\end{equation}%
all these eigenstates being regular on the definition domain.

The comparison of the expressions Eq(\ref{SIeigen}) and Eq(\ref{sawronsk1}),
allows to deduce directly Eq(\ref{wronskid}). 

\subsection{Extended Krein-Adler theorem and Wronskian identities
for mixed extensions of TSIP}

\bigskip By the above procedure, we have obtained an potential $V^{\left(
-M\right) }$ the spectrum of which being completely filled until the level $%
E_{-m}(\alpha )$. Up to a global translation and a shift in the parameters,
it coincides with the initial potential. By considering $V^{\left( -M\right)
}$ as an intermediate step, we can apply now an usual chain of
state-deleting DBT in the frame of the Krein-Adler theorem \cite{krein,adler}%
. It leads to suppress some sequences in the spectrum and the global chain
which is composed of both state-adding and state-deleting DBT is a \textit{%
mixed chain}.

We then obtain the following generalization of the Krein-Adler theorem

\begin{thm}[Extended Krein-Adler theorem for TSIP]

Consider a TSIP $V(x;\alpha ),\ \alpha \in U,$ with a
discrete spectrum $\left( E_{n},\psi _{n}\right) _{n\in \left\{
0,...,n_{\max }\right\} \mathbb{\subseteq N}}$ and a mixed chain of
state-adding and state-deleting DBT, characterized by the $m$-uple $N_{m}=\left( n_{1},...,n_{m}\right) $ of positive (state-deleting
DBT) and negative (state-adding DBT) integers. Let $n_{<}=\underset{1\leq
j\leq m}{\min }\left( n_{j}\right) $. When $\alpha _{j}\in U,\
\forall j\geq n_{<}$ the extended potential %
\begin{equation}
V^{\left( N_{m}\right) }(x;\alpha )=V(x;\alpha )+2\left( \ln \left\vert
W\left( \psi _{n_{1}},...,\psi _{n_{m}}\mid x;\alpha \right) \right\vert
\right) ^{\prime \prime }
\end{equation}%
is regular iff its spectrum is a subset of $\left\{ E_{j\in 
\mathbb{Z}}\right\} $ containing only even gaps (gaps constituted
by an even number of consecutive missing levels). 
\end{thm}

Before we address the proof, let us introduce the following notation:

Let $\pm L$ be the $l$-uple associated to a complete chain of $l$
state-adding DBT $-L=\left( -1,...,-l\right) $ or state-deleting DBT $%
L=\left( 0,1,...,l-1\right) $, $\left\{ l_{1},...,l_{k}\right\} $ a set of $%
k $ indices such that $\pm L\cap $ $\left\{ l_{1},...,l_{k}\right\} =\left\{
l_{i_{1}},...,l_{i_{j}}\right\} $ and $\left\{ l_{1},...,l_{k}\right\}
/\left\{ l_{i_{1}},...,l_{i_{j}}\right\} =\left\{
l_{p_{1}},...,l_{p_{k-j}}\right\} ,\quad j\leq l,k$. Then, $%
-L//l_{1},...,l_{k}$ is the $\left( l+k-2j\right) $-uple obtained by
suppressing in $\pm L$ the indices $l_{i_{1}},...,l_{i_{j}}$ and adding to
the resulting $\left( l-j\right) $-uple the indices $%
l_{p_{1}},...,l_{p_{k-j}}$

\begin{equation}
\pm L//l_{1},...,l_{k}=\left( \left( \pm L/l_{i_{1}},...,l_{i_{j}}\right)
,l_{p_{1}},...,l_{p_{k-j}}\right) .
\end{equation}

\textit{Proof.}
Let $-n_{<}=\underset{1\leq j\leq m}{\min }\left( n_{j}\right) $ and $%
-N_{<}=\left( -1,...,-n_{<}\right) $. We can always consider $N_{m}$ as
constituted by the juxtaposition of the complete chain $-N_{<}$ of $n_{<}$
state-adding DBT and of a chain $\left( n_{j_{1}},...,n_{j_{k}}\right) $ of $%
k$ state-deleting DBT such that $N_{m}=-N_{<}//n_{j_{1}},...,n_{j_{k}}$. $%
V^{\left( -N_{<}\right) }$ being regular (see Lemma \ref{lem4}), the Krein-Adler
theorem ensures that the extended potential generated by the chain of state
deleting DBT based on the eigenfunctions $\psi _{n_{j_{l}}}^{\left(
-N_{<}\right) }(x;\alpha ),\quad l=1,...,k,$ of $V^{\left( -N_{<}\right) }$
is regular iff the chain $\left( n_{j_{1}},...,n_{j_{k}}\right) $ is of the
Krein-Adler type. Using Lemma 2, the final extension $V^{\left(
-N_{<},n_{j_{1}},...,n_{j_{k}}\right) }$ can still be written

\begin{equation}
V^{\left( -N_{<},n_{j_{1}},...,n_{j_{k}}\right) }(x;\alpha )=V^{\left(
-N_{<}//n_{j_{1}},...,n_{j_{k}}\right) }(x;\alpha )=V^{\left( N_{m}\right)
}(x;\alpha ),
\end{equation}%
which is then regular iff its spectrum, that is, the set of integers $%
\left\{ n\geq -n_{<}\right\} /\left\{ n_{j_{1}},...,n_{j_{k}}\right\} $,
contains only even gaps. 

Combining the Lemmas \ref{lem2},\ref{lem4} and \ref{lem5} we obtain the following result

\begin{thm}\label{thm1}

Consider a mixed chain of DBT associated to $-M//n_{1},...,n_{k}$
(where $m<0$,$-M=\left( -1,...,-m\right) $, $-m\leq 
\underset{i=1,...,k}{\min }(n_{i})$ . Suppose that $\alpha
_{-j}\in U,\ \forall j\leq m$.Then, when this chain is applied to
the TSIP $V(x;\alpha )$,the final extension can be written as

\begin{equation}
V^{\left( -M//n_{1},...,n_{k}\right) }(x;\alpha )=V^{\left(
n_{1}+m,...,n_{k}+m\right) }(x;\alpha _{-m})+E_{-m}(\alpha ).  \label{Th1a}
\end{equation}%
and is regular iff $\left( n_{1},...,n_{k}\right) $ is
of the Krein-Adler type or equivalently iff the spectrum of $V^{\left(
-M//n_{1},...,n_{k}\right) }$ contains only even gaps. 

The eigenstates of $V^{\left( -M/n_{1},...,n_{k}\right) }$
can be expressed using  the following Wronskian representation for \mbox{$j\notin \left\{
n_{1},...,n_{k}\right\} ,\quad j\geq -m$ }
\begin{eqnarray}
\nonumber \psi _{j}^{\left( -M//n_{1},...,n_{k}\right) }(x;\alpha )&=&\frac{W^{\left(
-M//n_{1},...,n_{k,}j\right) }(x;\alpha )}{W^{\left(
-M//n_{1},...,n_{k}\right) }(x;\alpha )}\sim \frac{W^{\left(
n_{1}+m,...,n_{k}+m,j+m\right) }\left( x;\alpha _{-m}\right) }{W^{\left(
n_{1}+m,...,n_{k}+m\right) }\left( x;\alpha _{-m}\right) }\\
&=&\psi
_{j+m}^{\left( n_{1}+m,...,n_{k}+m\right) }(x;\alpha _{-m}).\quad
\label{Th1b}
\end{eqnarray}

\end{thm}

\textit{Proof.}
Using Eq(\ref{rsig}) and Eq(\ref{SIeigen}), we can write

\begin{eqnarray}
V^{\left( -M//n_{1},...,n_{k}\right) }(x;\alpha ) &=&V^{\left(
-M,n_{1},...,n_{k}\right) }(x;\alpha ) \\
&=&V^{\left( -M\right) }(x;\alpha )+2\left( \log \left( W\left( \psi
_{n_{1}}^{\left( -M\right) },...,\psi _{n_{k}}^{\left( -M\right) }\mid
x;\alpha \right) \right) \right) ^{\prime \prime }  \notag \\
&=&V(x;\alpha _{-m})+E_{-m}(\alpha )+2\left( \log \left( W\left( \psi
_{n_{1}+m},...,\psi _{n_{k}+m}\mid x;\alpha _{-m}\right) \right) \right)
^{\prime \prime }  \notag \\
&=&V^{\left( n_{1}+m,...,n_{k}+m\right) }(x;\alpha _{-m})+E_{-m}(\alpha ), 
\notag
\end{eqnarray}%
which implies

\begin{equation}
V^{\left( -M//n_{1},...,n_{k}\right) }(x;\alpha )=V^{\left(
n_{1}+m,...,n_{k}+m\right) }(x;\alpha _{-m})+E_{-m}(\alpha ).
\end{equation}

The regularity condition results from the extended Krein-Adler theorem for
TSIP.

Moreover, if $j\geq -m$

\begin{eqnarray}
\psi _{j}^{\left( -M//n_{1},...,n_{k}\right) }(x;\alpha ) &\sim &\psi
_{j}^{\left( -M,n_{1},...,n_{k}\right) }(x;\alpha )=\frac{W\left( \psi
_{n_{1}}^{\left( -M\right) },...,\psi _{n_{k}}^{\left( -M\right) },\psi
_{j}^{\left( -M\right) }\mid x;\alpha \right) }{W\left( \psi
_{n_{1}}^{\left( -M\right) },...,\psi _{n_{k}}^{\left( -M\right) }\mid
x;\alpha \right) } \\
&\sim &\frac{W\left( \psi _{n_{1}+m},...,\psi _{n_{k}+m},\psi _{j+m}\mid
x;\alpha _{-m}\right) }{W\left( \psi _{n_{1}+m},...,\psi _{n_{k}+m}\mid
x;\alpha _{-m}\right) }.  \notag
\end{eqnarray}

If $\alpha _{-m}\in U$, in which case $\psi _{j}^{\left(
n_{1}+m,...,n_{k}+m\right) }(x;\alpha _{-m})$ satisfies the required
Dirichlet boundary conditions, we then have

\begin{equation}
\psi _{j}^{\left( -M//n_{1},...,n_{k}\right) }(x;\alpha )\sim \psi
_{j+m}^{\left( n_{1}+m,...,n_{k}+m\right) }(x;\alpha _{-m}),
\end{equation}%
that is,

\begin{equation}
\frac{W^{\left( -M//n_{1},...,n_{k,}j\right) }(x;\alpha )}{W^{\left(
-M//n_{1},...,n_{k}\right) }(x;\alpha )}\sim \frac{W^{\left(
n_{1}+m,...,n_{k}+m,j+m\right) }\left( x;\alpha _{-m}\right) }{W^{\left(
n_{1}+m,...,n_{k}+m\right) }\left( x;\alpha _{-m}\right) }.
\end{equation}%

\section{ Examples }

In what follows, we apply the general formalism developed above to some
particular PTSIP. As an illustration, we consider the particular chain $%
-M//n_{1},...,n_{k}=\left( -3,1,2\right) $ for $m=3$, i.e. $-M=\left(
-1,-2,-3\right) $ and $\ k=4$ with $\left( n_{1},...,n_{4}\right) =\left(
-2,-1,1,2\right) $. Eq(\ref{Th1a}) and Eq(\ref{Th1a}) become in this case

\begin{equation}
\left\{ 
\begin{array}{c}
V^{\left( -3,1,2\right) }(x;\alpha )=V^{\left( 1,2,4,5\right) }(x;\alpha
_{-3})+E_{-3}(\alpha ) \\ 
W^{\left( -3,1,2,j\right) }(x;\alpha )/W^{\left( -3,1,2\right) }(x;\alpha
)\sim W^{\left( 1,2,4,5,j+3\right) }\left( x;\alpha _{-3}\right) /W^{\left(
1,2,4,5\right) }\left( x;\alpha _{-3}\right) .%
\end{array}%
\right.  \label{Th1Appl}
\end{equation}

\subsection{The harmonic oscillator}

The harmonic oscillator potential is defined on the real line by

\begin{equation}
V\left( x;\omega \right) =\frac{\omega ^{2}}{4}x^{2}-\frac{\omega }{2},\
\omega \in 
\mathbb{R}
.  \label{OH}
\end{equation}

With Dirichlet boundary conditions at infinity and supposing $\omega \in 
\mathbb{R}
^{+}$, $V\left( x;\omega \right) $ has the following spectrum ($z=\omega
x^{2}/2$)%
\begin{equation}
\left\{ 
\begin{array}{c}
E_{n}\left( \omega \right) =n\omega \\ 
\psi _{n}(x;\omega )=H_{n}\left( \sqrt{\omega /2}x\right) \exp \left(
-\omega x^{2}/4\right) =H_{n}\left( \sqrt{z}\right) \exp \left( -z/2\right)%
\end{array}%
\right. ,\quad n\geq 0.  \label{spec OH}
\end{equation}

It is the most simple example of TSIP, with $\alpha =\omega \in 
\mathbb{R}
$ and $\varepsilon =0$ (the parameter translation is of zero amplitude $%
\alpha _{1}=\alpha =\omega $), that is

\begin{equation}
V^{\left( 0\right) }\left( x;\omega \right) =V\left( x;\omega \right)
+\omega .
\end{equation}

The $\Gamma _{3}$ symmetry acts as \cite{grandati}

\begin{equation}
\omega \overset{\Gamma _{3}}{\rightarrow }\left( -\omega \right) ,\left\{ 
\begin{array}{c}
V(x;\omega )\overset{\Gamma _{3}}{\rightarrow }V(x;-\omega )=V(x;\omega
)+\omega \\ 
\psi _{n}(x;\omega )\overset{\Gamma _{3}}{\rightarrow }\psi _{n}(x;-\omega
)=\psi _{-(n+1)}(x;\omega ),%
\end{array}%
\right.
\end{equation}%
and then generates the following quasi-polynomial eigenfunctions of the
disconjugacy sector

\begin{equation}
\left\{ 
\begin{array}{c}
E_{-n}\left( \omega \right) =-n\omega \\ 
\psi _{-n}(x;\omega )=H_{n-1}\left( i\sqrt{\omega /2}x\right) \exp \left(
\omega x^{2}/4\right) =\psi _{n-1}(x;-\omega )%
\end{array}%
\right. ,\quad n\geq 1,  \label{L3hermite}
\end{equation}%
diverging at both infinities with a parity equal to the one of $n-1$. The
disconjugacy theorem implies immediately that $\psi _{-n}(x;\omega )$ is
free of node for $n$ odd. Then $1/\psi _{-n}$ is regular on the real line
and we can verify it is also normalizable. The DBT $A(w_{-n}\left( \omega
\right) ),$ $n$ odd, is consequently state-adding. Note that, for every odd $%
n$, the regular one-step extended potential $V^{\left( -n\right) }\left(
x;\omega \right) $ presents an even gap in its spectrum (the levels from $-1$
to $-n+1$ are missing), in agreement with the extended Krein-Adler theorem.
From Theorem \ref{thm1}, we deduce
\begin{equation}
V^{\left( -M//n_{1},...,n_{k}\right) }(x;\omega )=V(x;\omega )+2\left( \log
\left( W^{\left( -M//n_{1},...,n_{k}\right) }(x;\omega )\right) \right)
^{\prime \prime }=V^{\left( n_{1}+m,...,n_{k}+m\right) }(x;\omega )-m\omega
\end{equation}%
with any limitation on the length of the complete chain $-M$. Up to an
additive constant, the final potential coincides then exactly with the one
obtained by applying the chain of state-deleting DBT $\left(
n_{1},...,n_{k}\right) $ to the initial harmonic potential $V(x;\omega )$.
Moreover, we have
\begin{eqnarray}
\nonumber\psi _{j}^{\left( -M//n_{1},...,n_{k}\right) }(x;\omega )&=&\frac{W^{\left(
-M//n_{1},...,n_{k,}j\right) }(x;\omega )}{W^{\left(
-M//n_{1},...,n_{k}\right) }(x;\omega )}\sim \frac{W^{\left(
n_{1}+m,...,n_{k}+m,j+m\right) }\left( x;\omega \right) }{W^{\left(
n_{1}+m,...,n_{k}+m\right) }\left( x;\omega \right) }\\
&=&\psi _{j+m}^{\left(
n_{1}+m,...,n_{k}+m\right) }(x;\omega ),\quad j\geq -m.\quad
\end{eqnarray}
For the considered example(see Eq(\ref{Th1Appl})), we obtain
\begin{equation}
V^{\left( -3,1,2\right) }(x;\omega )=V^{\left( 1,2,4,5\right) }(x;\omega
)-3\omega .
\end{equation}%
and
\begin{equation}
\frac{W\left( \psi _{-3},\psi _{1},\psi _{2},\psi _{j}\mid x;\omega \right) 
}{W\left( \psi _{-3},\psi _{1},\psi _{2}\mid x;\omega \right) }\sim \frac{%
W\left( \psi _{1},\psi _{2},\psi _{4},\psi _{5},\psi _{j+3}\mid x;\omega
\right) }{W\left( \psi _{1},\psi _{2},\psi _{4},\psi _{5}\mid x;\omega
\right) }.
\end{equation}

Using Eq(\ref{spec OH}) and Eq(\ref{L3hermite}) and usual properties of
Wronskians \cite{muir}, this leads to ($\omega =2$)

\begin{equation}
\frac{W\left( H_{2}\left( ix\right) \exp \left( x^{2}\right) ,H_{1}\left(
x\right) ,H_{2}\left( x\right) ,H_{j}\left( x\right) \mid x\right) }{W\left(
H_{2}\left( ix\right) \exp \left( x^{2}\right) ,H_{1}\left( x\right)
,H_{2}\left( x\right) \mid x\right) }\sim \frac{W\left( H_{1}\left( x\right)
,H_{2}\left( x\right) ,H_{4}\left( x\right) ,H_{5}\left( x\right)
,H_{j+3}\left( x\right) \mid x\right) }{W\left( H_{1}\left( x\right)
,H_{2}\left( x\right) ,H_{4}\left( x\right) ,H_{5}\left( x\right) \mid
x\right) }.
\end{equation}

From the Rodrigues formula \cite{szego,magnus}, it results

\begin{equation}
\frac{d^{k}}{dx^{k}}\left( \exp \left( x^{2}\right) H_{n}\left( ix\right)
\right) =\frac{1}{i^{k}}\exp \left( x^{2}\right) H_{n+k}\left( ix\right) ,
\end{equation}%
and since 
\begin{equation}
\frac{d^{k}}{dx^{k}}\left( H_{n}\left( x\right) \right) =2^{k}\frac{n!}{%
(n-k)!}H_{n-k}\left( x\right) ,
\end{equation}%
we obtain the following bilinear determinantal identity for Hermite
polynomials

\begin{equation}
\small
\frac{\left\vert 
\begin{array}{cccc}
H_{2}\left( ix\right) & H_{1}\left( x\right) & H_{2}\left( x\right) & 
H_{j}\left( x\right) \\ 
\frac{1}{i}H_{3}\left( ix\right) & 2H_{0}\left( x\right) & 4H_{1}\left(
x\right) & 2jH_{j-1}\left( x\right) \\ 
-H_{4}\left( ix\right) & 0 & 8H_{0}\left( x\right) & 4(j-1)_{2}H_{j-2}\left(
x\right) \\ 
-\frac{1}{i}H_{5}\left( ix\right) & 0 & 0 & 8\left( j-2\right)
_{3}H_{j-3}\left( x\right)%
\end{array}%
\right\vert }{\left\vert 
\begin{array}{ccc}
H_{2}\left( ix\right) & H_{1}\left( x\right) & H_{2}\left( x\right) \\ 
\frac{1}{i}H_{3}\left( ix\right) & 2H_{0}\left( x\right) & 4H_{1}\left(
x\right) \\ 
-H_{4}\left( ix\right) & 0 & 8H_{0}\left( x\right)%
\end{array}%
\right\vert }\sim \frac{\left\vert 
\begin{array}{ccccc}
H_{1}\left( x\right) & H_{2}\left( x\right) & H_{4}\left( x\right) & 
H_{5}\left( x\right) & H_{j+3}\left( x\right) \\ 
H_{0}\left( x\right) & 2H_{1}\left( x\right) & 4H_{3}\left( x\right) & 
5H_{4}\left( x\right) & (j+3)H_{j+2}\left( x\right) \\ 
0 & 2H_{0}\left( x\right) & 12H_{2}\left( x\right) & 20H_{3}\left( x\right)
& (j+2)_{2}H_{j+1}\left( x\right) \\ 
0 & 0 & 24H_{1}\left( x\right) & 60H_{2}\left( x\right) & 
(j+1)_{3}H_{j}\left( x\right) \\ 
0 & 0 & 24H_{0}\left( x\right) & 120H_{1}\left( x\right) & \left( j\right)
_{4}H_{j-1}\left( x\right)%
\end{array}%
\right\vert }{\left\vert 
\begin{array}{cccc}
H_{1}\left( x\right) & H_{2}\left( x\right) & H_{4}\left( x\right) & 
H_{5}\left( x\right) \\ 
H_{0}\left( x\right) & 2H_{1}\left( x\right) & 4H_{3}\left( x\right) & 
5H_{4}\left( x\right) \\ 
0 & 2H_{0}\left( x\right) & 12H_{2}\left( x\right) & 20H_{3}\left( x\right)
\\ 
0 & 0 & 24H_{1}\left( x\right) & 60H_{2}\left( x\right)%
\end{array}%
\right\vert },
\end{equation}%
where $\left( x\right) _{n}$ is the usual Pochammer symbol $\left( x\right) _{n}=x(x+1)...(x+n-1)$.

\subsection{The isotonic oscillator}

The isotonic oscillator potential (with zero ground level $E_{0}=0$)) is
defined on the positive half line $\left] 0,+\infty \right[ $ by

\begin{equation}
V\left( x;\omega ,\alpha \right) =\frac{\omega ^{2}}{4}x^{2}+\frac{\left(
\alpha +1/2\right) (\alpha -1/2)}{x^{2}}-\omega \left( \alpha +1\right)
,\quad \left\vert \alpha \right\vert >1/2.  \label{OI}
\end{equation}

If we add Dirichlet boundary conditions at $0$ and infinity and if we
suppose $\alpha >1/2$ and $\alpha \notin 
\mathbb{N}
$, it has the following spectrum ($z=\omega x^{2}/2$)%
\begin{equation}
\left\{ 
\begin{array}{c}
E_{n}\left( \omega \right) =2n\omega \\ 
\psi _{n}\left( x;\omega ,\alpha \right) =x^{\alpha +1/2}e^{-z/2}\mathit{L}%
_{n}^{\alpha }\left( z\right)%
\end{array}%
\right. ,\quad n\geq 0.  \label{spec OI}
\end{equation}

It is a TSIP, with $\left( \omega ,\alpha \right) \in 
\mathbb{R}
^{2}$ and $\varepsilon =\left( 0,+1\right) $

\begin{equation}
V^{\left( 0\right) }\left( x;\omega ,\alpha \right) =V\left( x;\omega
,\alpha _{1}\right) +2\omega .
\end{equation}

It possesses a $\Gamma _{3}$ symmetry which acts as \cite{grandati}

\begin{equation}
\left( \omega ,\alpha \right) \overset{\Gamma _{3}}{\rightarrow }\left(
-\omega ,-\alpha \right) ,\left\{ 
\begin{array}{c}
V(x;\omega ,\alpha )\overset{\Gamma _{3}}{\rightarrow }V(x;-\omega ,-\alpha
)=V(x;\omega ,\alpha )+\omega \\ 
\psi _{n}(x;\omega ,\alpha )\overset{\Gamma _{3}}{\rightarrow }\psi
_{n}(x;-\omega ,-\alpha )=\psi _{-(n+1)}(x;\omega ,\alpha ),%
\end{array}%
\right.
\end{equation}%
and then generates the following quasi-polynomial eigenfunctions of the
disconjugacy sector

\begin{equation}
\left\{ 
\begin{array}{c}
E_{-n}\left( \omega \right) =-2n\omega \\ 
\psi _{-n}(x;\omega ,\alpha )=x^{-\alpha +1/2}e^{z/2}\mathit{L}%
_{n-1}^{-\alpha }\left( -z\right)%
\end{array}%
\right. ,\quad n\geq 1.  \label{L3Lag}
\end{equation}

Since \cite{szego,magnus}%
\begin{equation}
\left\{ 
\begin{array}{c}
\mathit{L}_{n}^{-\alpha }\left( -x\right) \underset{x\rightarrow 0^{+}}{%
\rightarrow }\frac{\left( -\alpha +1\right) _{n}}{n!}=\frac{\left( -\alpha
+1\right) ...\left( -\alpha +n\right) }{n!} \\ 
\mathit{L}_{n}^{-\alpha }\left( -x\right) \underset{x\rightarrow +\infty }{%
\sim }\frac{1}{n!}x^{n},%
\end{array}%
\right.  \label{asymptLag}
\end{equation}%
$\psi _{-n}$ diverges at $0$ and infinity ($\alpha >1/2$). Moreover, when $%
n<\alpha +1$, $\psi _{-n}$ has an asymptotic sign equal to $\left( -1\right)
^{n-1}$ and when $n>1+\left[ \alpha \right] $ ($\left[ \alpha \right] $
being the integer part of $\alpha $), $\psi _{-n}$ has an asymptotic sign
equal to $\left( -1\right) ^{\left[ \alpha \right] }$, independently of $n$.

\bigskip Note that the disconjugacy theorem implies immediately that for $%
n<\alpha +1$, $\psi _{-n}(x;\omega ,\alpha )$ is free of node when $n$ is
odd and that for $n>1+\left[ \alpha \right] $, $\psi _{-n}(x;\omega ,\alpha
) $ is free of node when $\left[ \alpha \right] $ is even. Then $1/\psi
_{-n} $ is regular on the real line and it is also normalizable (see Eq(\ref%
{L3Lag})). This is in agreement with the classical Kienast-Lawton-Hahn
theorem on the number of negative zeros of the Laguerre polynomials \cite%
{szego,magnus,grandati}. Consequently, in these cases, the DBT $%
A(w_{-n}\left( \omega ,\alpha \right) )$ is state-adding.

When $n<\alpha +1$ and $n$ odd, the regular one-step extended potential $%
V^{\left( -n\right) }\left( x;\omega ,\alpha \right) $ presents an even gap
in its spectrum (the levels from $-1$ to $-n+1$ are missing), in agreement
with the extended Krein-Adler theorem.
However, when $n>1+\left[ \alpha \right] $ and $\left[ \alpha \right] $
even, $V^{\left( -n\right) }\left( x;\omega ,\alpha \right) $ is regular for
any parity of $n$ and consequently its spectrum can presents odd gaps (when $%
n$ is even, the $n-1$ levels $-n-1,...,-1$ are missing). Nevertheless, for $%
m>1+\left[ \alpha \right] $ we satisfy no longer the condition $\left\vert
\alpha _{-j}\right\vert <1/2,\ \forall j\leq m$. Indeed, $\left\vert \alpha
_{-\left[ \alpha \right] }\right\vert >1/2$ or $\left\vert \alpha _{-\left( %
\left[ \alpha \right] +1\right) }\right\vert >1/2$ and the potentials $%
V\left( x;\omega ,\alpha _{-\left[ \alpha \right] }\right) $ or $V\left(
x;\omega ,\alpha _{-\left( \left[ \alpha \right] +1\right) }\right) $ have a
singular behaviour.

To stay in the conditions of applicability of the extended Krein-Adler
theorem and Theorem \ref{thm1}, we have  to consider only seed functions
satisfying $n<\alpha +1$.

If $\alpha >m+1/2$, we have (see Theorem \ref{thm1})

\begin{equation}
V^{\left( -M//n_{1},...,n_{k}\right) }(x;\omega ,\alpha )=V^{\left(
n_{1}+m,...,n_{k}+m\right) }(x;\omega ,\alpha _{-m})-2m\omega .
\end{equation}

The eigenstates of this extended potential satisfy for $j\geq -m$

\begin{eqnarray}
\qquad \nonumber\psi _{j}^{\left( -M//n_{1},...,n_{k}\right) }(x;\omega ,\alpha )&=&\frac{%
W^{\left( -M//n_{1},...,n_{k,}j\right) }(x;\omega ,\alpha )}{W^{\left(
-M//n_{1},...,n_{k}\right) }(x;\omega ,\alpha )}\sim \frac{W^{\left(
n_{1}+m,...,n_{k}+m,j+m\right) }\left( x;\omega ,\alpha _{-m}\right) }{%
W^{\left( n_{1}+m,...,n_{k}+m\right) }\left( x;\omega ,\alpha _{-m}\right) }%
\\
&=&\psi _{j+m}^{\left( n_{1}+m,...,n_{k}+m\right) }(x;\omega ,\alpha
_{-m}).\quad
\end{eqnarray}

In the considered example we obtain, for $\alpha >7/2$

\begin{equation}
V^{\left( -3,1,2\right) }(x;\omega ,\alpha )=V^{\left( 1,2,4,5\right)
}(x;\omega ,\alpha _{-3})-3\omega .
\end{equation}%
and

\begin{equation}
\frac{W\left( \psi _{-3},\psi _{1},\psi _{2},\psi _{j}\mid x;\omega ,\alpha
\right) }{W\left( \psi _{-3},\psi _{1},\psi _{2}\mid x;\omega ,\alpha
\right) }\sim \frac{W\left( \psi _{1},\psi _{2},\psi _{4},\psi _{5},\psi
_{j+3}\mid x;\omega ,\alpha _{-3}\right) }{W\left( \psi _{1},\psi _{2},\psi
_{4},\psi _{5}\mid x;\omega ,\alpha _{-3}\right) }.
\end{equation}

Using Eq(\ref{spec OI}) and Eq(\ref{L3Lag}) and the usual properties of
Wronskians \cite{muir}, this leads to

\begin{equation}
\frac{W\left( z^{-\alpha }e^{z}\mathit{L}_{2}^{-\alpha }\left( -z\right) ,%
\mathit{L}_{1}^{\alpha }\left( z\right) ,\mathit{L}_{2}^{\alpha }\left(
z\right) ,\mathit{L}_{j}^{\alpha }\left( z\right) \mid x\right) }{W\left(
z^{-\alpha }e^{z}\mathit{L}_{2}^{-\alpha }\left( -z\right) ,\mathit{L}%
_{1}^{\alpha }\left( z\right) ,\mathit{L}_{2}^{\alpha }\left( z\right) \mid
x\right) }\sim \frac{W\left( \mathit{L}_{1}^{\alpha _{-3}}\left( z\right) ,%
\mathit{L}_{2}^{\alpha _{-3}}\left( z\right) ,\mathit{L}_{4}^{\alpha
_{-3}}\left( z\right) ,\mathit{L}_{5}^{\alpha _{-3}}\left( z\right) ,\mathit{%
L}_{j+3}^{\alpha _{-3}}\left( z\right) \mid x\right) }{W\left( \mathit{L}%
_{1}^{\alpha _{-3}}\left( z\right) ,\mathit{L}_{2}^{\alpha _{-3}}\left(
z\right) ,\mathit{L}_{4}^{\alpha _{-3}}\left( z\right) ,\mathit{L}%
_{5}^{\alpha _{-3}}\left( z\right) \mid x\right) },
\end{equation}%
which can be also expressed, using the formula for the change of variable $$W\left( y_{1},...,y_{m}\mid x\right) =\left( \frac{dz}{dx}\right)
^{m(m-1)/2}W\left( y_{1},...,y_{m}\mid z\right), $$  as

\begin{equation}
z\frac{W\left( z^{-\alpha }e^{z}\mathit{L}_{2}^{-\alpha }\left( -z\right) ,%
\mathit{L}_{1}^{\alpha }\left( z\right) ,\mathit{L}_{2}^{\alpha }\left(
z\right) ,\mathit{L}_{j}^{\alpha }\left( z\right) \mid z\right) }{W\left(
z^{-\alpha }e^{z}\mathit{L}_{2}^{-\alpha }\left( -z\right) ,\mathit{L}%
_{1}^{\alpha }\left( z\right) ,\mathit{L}_{2}^{\alpha }\left( z\right) \mid
z\right) }\sim \frac{W\left( \mathit{L}_{1}^{\alpha _{-3}}\left( z\right) ,%
\mathit{L}_{2}^{\alpha _{-3}}\left( z\right) ,\mathit{L}_{4}^{\alpha
_{-3}}\left( z\right) ,\mathit{L}_{5}^{\alpha _{-3}}\left( z\right) ,\mathit{%
L}_{j+3}^{\alpha _{-3}}\left( z\right) \mid z\right) }{W\left( \mathit{L}%
_{1}^{\alpha _{-3}}\left( z\right) ,\mathit{L}_{2}^{\alpha _{-3}}\left(
z\right) ,\mathit{L}_{4}^{\alpha _{-3}}\left( z\right) ,\mathit{L}%
_{5}^{\alpha _{-3}}\left( z\right) \mid z\right) }.  \label{wronsklag}
\end{equation}

The following identity for Laguerre polynomials \cite{szego,magnus}

\begin{equation}
\frac{d^{k}}{dz^{k}}\left( \mathit{L}_{n}^{\alpha }\left( z\right) \right)
=\left( -1\right) ^{k}\mathit{L}_{n-k}^{\alpha +k}\left( z\right)
\end{equation}%
can be used in the Rodrigues formula for Laguerre polynomials to give
\begin{equation}
\frac{d^{k}}{dz^{k}}\left( z^{-\alpha }e^{z}\mathit{L}_{n}^{-\alpha }\left(
-z\right) \right) =\frac{\left( n+k\right) !}{n!}z^{-\alpha -k}e^{z}\mathit{L%
}_{n+k}^{-\alpha -k}\left( -z\right) .
\end{equation}

Eq(\ref{wronsklag}) then gives the following bilinear determinantal identity
for Laguerre polynomials
\begin{equation}
z\,\small\frac{\left\vert 
\begin{array}{cccc}
\mathit{L}_{2}^{-\alpha }\left( -z\right) & \mathit{L}_{1}^{\alpha }\left(
z\right) & \mathit{L}_{2}^{\alpha }\left( z\right) & \mathit{L}_{j}^{\alpha
}\left( z\right) \\ 
-3z^{-1}\mathit{L}_{3}^{-\alpha -1}\left( -z\right) & \mathit{L}_{0}^{\alpha
+1}\left( z\right) & \mathit{L}_{1}^{\alpha +1}\left( z\right) & \mathit{L}%
_{j-1}^{\alpha +1}\left( z\right) \\ 
12z^{-2}\mathit{L}_{4}^{-\alpha -2}\left( -z\right) & 0 & \mathit{L}%
_{0}^{\alpha +2}\left( z\right) & \mathit{L}_{j-2}^{\alpha +2}\left( z\right)
\\ 
-60z^{-3}\mathit{L}_{5}^{-\alpha -3}\left( -z\right) & 0 & 0 & \mathit{L}%
_{j-3}^{\alpha +3}\left( z\right)%
\end{array}%
\right\vert }{\left\vert 
\begin{array}{ccc}
\mathit{L}_{2}^{-\alpha }\left( -z\right) & \mathit{L}_{1}^{\alpha }\left(
z\right) & \mathit{L}_{2}^{\alpha }\left( z\right) \\ 
-3z^{-1}\mathit{L}_{3}^{-\alpha -1}\left( -z\right) & \mathit{L}_{0}^{\alpha
+1}\left( z\right) & \mathit{L}_{1}^{\alpha +1}\left( z\right) \\ 
12z^{-2}\mathit{L}_{4}^{-\alpha -2}\left( -z\right) & 0 & \mathit{L}%
_{0}^{\alpha +2}\left( z\right)%
\end{array}%
\right\vert }\sim \frac{\left\vert 
\begin{array}{ccccc}
\mathit{L}_{1}^{\alpha -3}\left( z\right) & \mathit{L}_{2}^{\alpha -3}\left(
z\right) & \mathit{L}_{4}^{\alpha -3}\left( z\right) & \mathit{L}%
_{5}^{\alpha -3}\left( z\right) & \mathit{L}_{j+3}^{\alpha -3}\left( z\right)
\\ 
\mathit{L}_{0}^{\alpha -2}\left( z\right) & \mathit{L}_{1}^{\alpha -2}\left(
z\right) & \mathit{L}_{3}^{\alpha -2}\left( z\right) & \mathit{L}%
_{4}^{\alpha -2}\left( z\right) & \mathit{L}_{j+2}^{\alpha -2}\left( z\right)
\\ 
0 & \mathit{L}_{0}^{\alpha -1}\left( z\right) & \mathit{L}_{2}^{\alpha
-1}\left( z\right) & \mathit{L}_{3}^{\alpha -1}\left( z\right) & \mathit{L}%
_{j+1}^{\alpha -1}\left( z\right) \\ 
0 & 0 & \mathit{L}_{1}^{\alpha }\left( z\right) & \mathit{L}_{2}^{\alpha
}\left( z\right) & \mathit{L}_{j}^{\alpha }\left( z\right) \\ 
0 & 0 & \mathit{L}_{0}^{\alpha +1}\left( z\right) & \mathit{L}_{1}^{\alpha
+1}\left( z\right) & \mathit{L}_{j-1}^{\alpha +1}\left( z\right)%
\end{array}%
\right\vert }{\left\vert 
\begin{array}{cccc}
\mathit{L}_{1}^{\alpha -3}\left( z\right) & \mathit{L}_{2}^{\alpha -3}\left(
z\right) & \mathit{L}_{4}^{\alpha -3}\left( z\right) & \mathit{L}%
_{5}^{\alpha -3}\left( z\right) \\ 
\mathit{L}_{0}^{\alpha -2}\left( z\right) & \mathit{L}_{1}^{\alpha -2}\left(
z\right) & \mathit{L}_{3}^{\alpha -2}\left( z\right) & \mathit{L}%
_{4}^{\alpha -2}\left( z\right) \\ 
0 & \mathit{L}_{0}^{\alpha -1}\left( z\right) & \mathit{L}_{2}^{\alpha
-1}\left( z\right) & \mathit{L}_{3}^{\alpha -1}\left( z\right) \\ 
0 & 0 & \mathit{L}_{1}^{\alpha }\left( z\right) & \mathit{L}_{2}^{\alpha
}\left( z\right)%
\end{array}%
\right\vert }.
\end{equation}

\subsection{The trigonometric Darboux-P\"{o}schl-Teller (TDPT) potential}

The trigonometric Darboux-P\"{o}schl-Teller (TDPT) potential is defined on
the interval $\left] 0,\pi /2\right[ $ by

\begin{equation}
V(x;\alpha ,\beta )=\frac{(\alpha +1/2)(\alpha -1/2)}{\cos ^{2}x}+\frac{%
(\beta +1/2)(\beta -1/2)}{\sin ^{2}x}-(\alpha +\beta +1)^{2},\ \left\vert
\alpha \right\vert ,\left\vert \beta \right\vert >1/2.  \label{TDPT}
\end{equation}

With Dirichlet boundary conditions at $0$ and $\pi /2$, it has, in the case $%
\alpha ,\beta >1/2$, the following spectrum 
\begin{equation}
\left\{ 
\begin{array}{c}
\text{ }E_{n}\left( \alpha ,\beta \right) =(\alpha _{n}+\beta
_{n}+1)^{2}-(\alpha +\beta +1)^{2}=4n(\alpha +\beta +1+n) \\ 
\\ 
\psi _{n}\left( x;\alpha ,\beta \right) =\left( \sin x\right) ^{\alpha
+1/2}\left( \cos x\right) ^{\beta +1/2}\mathit{P}_{n}^{\left( \alpha ,\beta
\right) }\left( \cos 2x\right)%
\end{array}%
\right. ,\ n\in \mathbb{N},  \label{spec TDPT}
\end{equation}%
where $\mathit{P}_{n}^{\left( \alpha ,\beta \right) }$ are the usual Jacobi
polynomials \cite{szego,magnus} and where $(\alpha _{n},\beta _{n})=(\alpha
+n,\beta +n)$.

It is a TSIP, with $\left( \alpha ,\beta \right) \in 
\mathbb{R}
^{2}$ and $\varepsilon =\left( +1,+1\right) $

\begin{equation}
V^{\left( 0\right) }\left( x;\alpha ,\beta \right) =V\left( x;\alpha
_{1},\beta _{1}\right) +4(\alpha +\beta +2).
\end{equation}

For this potential, the $\Gamma _{3}$ symmetry acts as

\begin{equation}
\left( \alpha ,\beta \right) \overset{\Gamma _{3}}{\rightarrow }\left(
-\alpha ,-\beta \right) ,\left\{ 
\begin{array}{c}
V(x;\alpha ,\beta )\overset{\Gamma _{3}}{\rightarrow }V(x;-\alpha ,-\beta
)=V(x;\alpha ,\beta )+4\alpha (\beta +1) \\ 
\psi _{n}(x;\alpha ,\beta )\overset{\Gamma _{3}}{\rightarrow }\psi
_{n}(x;-\alpha ,-\beta ),%
\end{array}%
\right.  \label{gamma3DPT}
\end{equation}%
where $\psi _{n}(x;-\alpha ,-\beta )$ satisfies

\begin{equation}
\widehat{H}(-\alpha ,-\beta )\psi _{n}(x;-\alpha ,-\beta )=E_{n}\left(
-\alpha ,-\beta \right) \psi _{n}(x;-\alpha ,-\beta ),
\end{equation}%
that is, using Eq(\ref{gamma3DPT})

\begin{eqnarray}
\widehat{H}(\alpha ,\beta )\psi _{n}(x;-\alpha ,-\beta ) &=&\left(
4n(1+n-\alpha -\beta )-4\alpha (\beta +1)\right) \psi _{n}(x;-\alpha ,-\beta
) \\
&=&E_{-(n+1)}\left( \alpha ,\beta \right) \psi _{n}(x;-\alpha ,-\beta ), 
\notag
\end{eqnarray}%
from which we deduce $\mathcal{E}_{n,3}(\alpha ,\beta )=E_{-(n+1)}\left(
\alpha ,\beta \right) $ and

\begin{equation}
\psi _{-(n+1)}(x;\alpha ,\beta )=\psi _{n}(x;-\alpha ,-\beta ).
\end{equation}

Then $\Gamma _{3}$ generates the following quasi-polynomial eigenfunctions
of $V\left( x;\alpha ,\beta \right) $.

\begin{equation}
\left\{ 
\begin{array}{c}
E_{-n}\left( \alpha ,\beta \right) =4n(n-1-\alpha -\beta ) \\ 
\psi _{-n}(x;\omega ,\alpha )=\left( \sin x\right) ^{-\alpha +1/2}\left(
\cos x\right) ^{-\beta +1/2}\mathit{P}_{n-1}^{\left( -\alpha ,-\beta \right)
}\left( \cos 2x\right) .%
\end{array}%
\right.  \label{J3}
\end{equation}

In this case they are only a finite number of them (satisfying $n<1+\alpha
+\beta $ in which case $E_{-n}\left( \alpha ,\beta \right) <0$) which are in
the disconjugacy sector. Note that the quadratic dispersion relation ($%
E_{n}\left( \alpha ,\beta \right) $ as a function of $n$) is symmetric with
respect to the value $-\frac{1+\alpha +\beta }{2}$ implying that the values
of $-n$ we have to take into account are in fact limited to the set $\left\{
-\left[ \left( 1+\alpha +\beta \right) /2\right] ,...,-1\right\} $ ($\left[
\lambda \right] $ being the integer part of $\lambda $). This corresponds to
the fact that \cite{szego,magnus}

\begin{equation}
\mathit{P}_{n}^{\left( -\alpha ,-\beta \right) }\left( z\right) \sim F\left(
-n,n+1+\alpha +\beta ,\alpha +1,\frac{1-x}{2}\right) ,
\end{equation}%
the right hand member being invariant under the transformation $n\rightarrow
-\left( n+1+\alpha +\beta \right) $, i.e. the symmetry of $n$ with respect to $%
-\frac{1+\alpha +\beta }{2}$.

$\psi _{-n}$ diverges at $0$ and infinity ($\alpha ,\beta >1/2$). Moreover 
\cite{szego,magnus}%
\begin{equation}
\left\{ 
\begin{array}{c}
\mathit{P}_{n}^{\left( -\alpha ,-\beta \right) }\left( z\right) \underset{%
z\rightarrow 1^{-}}{\rightarrow }\frac{\left( -\alpha +1\right) _{n}}{n!}=%
\frac{\left( -\alpha +1\right) ...\left( -\alpha +n\right) }{n!} \\ 
\mathit{P}_{n}^{\left( -\alpha ,-\beta \right) }\left( z\right) \underset{%
z\rightarrow -1+}{\rightarrow }\left( -1\right) ^{n}\frac{\left( -\beta
+1\right) _{n}}{n!}=\frac{\left( \beta -1\right) ...\left( \beta -n\right) }{%
n!}.%
\end{array}%
\right.  \label{asymptJac}
\end{equation}

When $\left( -\alpha +1\right) _{n-1}$ and $\left( -1\right) ^{n-1}\left(
-\beta +1\right) _{n-1}$ have the same sign or have opposite signs, $\psi
_{-n}$ has an even or odd asymptotic parity. If we restrict ourselves to $%
n<\alpha +1$ and $n<\beta +1$, in which case we also have immediately $%
n<\left( 1+\alpha +\beta \right) /2$, the disconjugacy theorem then implies
that $\psi _{-n}$ is free of node when $n$ is odd. Moreover this is the
condition to satisfy in order that $\left\vert \alpha _{-j}\right\vert
,\left\vert \beta _{-j}\right\vert <1/2,\ \forall j\leq n,$ which, as in the
example of the isotonic oscillator, is a necessary condition for the use of
the extended Krein-Adler theorem and of Theorem \ref{thm1}, to chains of DBT
including seed functions until $\psi _{-n}$. We have then to consider only
seed functions satisfying $n<\min \left( \alpha +1,\beta +1\right) $.

\bigskip Note that the disconjugacy theorem implies immediately that in this
case for $n$ odd, $\psi _{-n}(x;\alpha ,\beta )$ is free of node. Then $%
1/\psi _{-n}$ is regular on the real line, normalizable (see Eq(\ref{L3Lag}%
)) and the DBT $A(w_{-n}\left( \alpha ,\beta \right) )$ is state-adding.

From Theorem \ref{thm1}, if $\alpha ,\beta >m+1/2$, we then obtain a extended
potential of the form
\begin{equation}
V^{\left( -M//n_{1},...,n_{k}\right) }(x;\alpha ,\beta )=V^{\left(
n_{1}+m,...,n_{k}+m\right) }(x;\alpha _{-m},\beta _{-m})-4m(1+\alpha +\beta
-m),
\end{equation}%
the eigenstates of which satisfy, for $j\geq -m$

\begin{eqnarray}
\nonumber\psi _{j}^{\left( -M//n_{1},...,n_{k}\right) }(x;\alpha ,\beta )&=&\frac{%
W^{\left( -M//n_{1},...,n_{k,}j\right) }(x;\alpha ,\beta )}{W^{\left(
-M//n_{1},...,n_{k}\right) }(x;\alpha ,\beta )}\sim \frac{W^{\left(
n_{1}+m,...,n_{k}+m,j+m\right) }\left( x;\alpha _{-m},\beta _{-m}\right) }{%
W^{\left( n_{1}+m,...,n_{k}+m\right) }\left( x;\alpha _{-m},\beta
_{-m}\right) }\\
&=&\psi _{j+m}^{\left( n_{1}+m,...,n_{k}+m\right) }(x;\alpha
_{-m},\beta _{-m}).\quad
\end{eqnarray}

In the example considered above (see Eq(\ref{Th1Appl})), we obtain for $%
\alpha ,\beta >7/2$

\begin{equation}
V^{\left( -3,1,2\right) }(x;\alpha ,\beta )=V^{\left( 1,2,4,5\right)
}(x;\alpha ,\beta )-12(\alpha +\beta -2).
\end{equation}%
and%
\begin{equation}
\frac{W\left( \psi _{-3},\psi _{1},\psi _{2},\psi _{j}\mid x;\alpha ,\beta
\right) }{W\left( \psi _{-3},\psi _{1},\psi _{2}\mid x;\alpha ,\beta \right) 
}\sim \frac{W\left( \psi _{1},\psi _{2},\psi _{4},\psi _{5},\psi _{j+3}\mid
x;\alpha _{-3},\beta _{-3}\right) }{W\left( \psi _{1},\psi _{2},\psi
_{4},\psi _{5}\mid x;\alpha _{-3},\beta _{-3}\right) },
\end{equation}

Using Eq(\ref{spec TDPT}) and Eq(\ref{J3}) and usual properties of
Wronskians \cite{muir}, this leads to ($z=\cos 2x$)

\begin{eqnarray}
&&\frac{W\left( \left( \sin x\right) ^{-2\alpha }\left( \cos x\right)
^{-2\beta }\mathit{P}_{2}^{\left( -\alpha ,-\beta \right) }\left( z\right) ,%
\mathit{P}_{1}^{\left( \alpha ,\beta \right) }\left( z\right) ,\mathit{P}%
_{2}^{\left( \alpha ,\beta \right) }\left( z\right) ,\mathit{P}_{j}^{\left(
\alpha ,\beta \right) }\left( z\right) \mid x\right) }{W\left( \left( \sin
x\right) ^{-2\alpha }\left( \cos x\right) ^{-2\beta }\mathit{P}_{2}^{\left(
-\alpha ,-\beta \right) }\left( z\right) ,\mathit{P}_{1}^{\left( \alpha
,\beta \right) }\left( z\right) ,\mathit{P}_{2}^{\left( \alpha ,\beta
\right) }\left( z\right) \mid x\right) } \\
&\sim &\left( \sin 2x\right) ^{-3}\frac{W\left( \mathit{P}_{1}^{\left(
\alpha _{-3},\beta _{-3}\right) }\left( z\right) ,\mathit{P}_{1}^{\left(
\alpha _{-3},\beta _{-3}\right) }\left( z\right) ,\mathit{P}_{1}^{\left(
\alpha _{-3},\beta _{-3}\right) }\left( z\right) ,\mathit{P}_{5}^{\left(
\alpha _{-3},\beta _{-3}\right) }\left( z\right) ,\mathit{P}_{j+3}^{\left(
\alpha _{-3},\beta _{-3}\right) }\left( z\right) \mid x\right) }{W\left( 
\mathit{P}_{1}^{\left( \alpha _{-3},\beta _{-3}\right) }\left( z\right) ,%
\mathit{P}_{1}^{\left( \alpha _{-3},\beta _{-3}\right) }\left( z\right) ,%
\mathit{P}_{1}^{\left( \alpha _{-3},\beta _{-3}\right) }\left( z\right) ,%
\mathit{P}_{5}^{\left( \alpha _{-3},\beta _{-3}\right) }\left( z\right) \mid
x\right) },  \notag
\end{eqnarray}%
or, with $W\left( y_{1},...,y_{m}\mid x\right) =\left( \frac{dz}{dx}\right)
^{m(m-1)/2}W\left( y_{1},...,y_{m}\mid z\right) $ \cite{muir}

\begin{eqnarray}
&&\left( 1-z^{2}\right) \frac{W\left( \left( 1-z\right) ^{-\alpha }\left(
1+z\right) ^{-\beta }\mathit{P}_{2}^{\left( -\alpha ,-\beta \right) }\left(
z\right) ,\mathit{P}_{1}^{\left( \alpha ,\beta \right) }\left( z\right) ,%
\mathit{P}_{2}^{\left( \alpha ,\beta \right) }\left( z\right) ,\mathit{P}%
_{j}^{\left( \alpha ,\beta \right) }\left( z\right) \mid z\right) }{W\left(
\left( 1-z\right) ^{-\alpha }\left( 1+z\right) ^{-\beta }\mathit{P}%
_{2}^{\left( -\alpha ,-\beta \right) }\left( z\right) ,\mathit{P}%
_{1}^{\left( \alpha ,\beta \right) }\left( z\right) ,\mathit{P}_{2}^{\left(
\alpha ,\beta \right) }\left( z\right) \mid z\right) }  \label{wronskjacobi}
\\
&\sim &\frac{W\left( \mathit{P}_{1}^{\left( \alpha _{-3},\beta _{-3}\right)
}\left( z\right) ,\mathit{P}_{2}^{\left( \alpha _{-3},\beta _{-3}\right)
}\left( z\right) ,\mathit{P}_{4}^{\left( \alpha _{-3},\beta _{-3}\right)
}\left( z\right) ,\mathit{P}_{5}^{\left( \alpha _{-3},\beta _{-3}\right)
}\left( z\right) ,\mathit{P}_{j+3}^{\left( \alpha _{-3},\beta _{-3}\right)
}\left( z\right) \mid z\right) }{W\left( \mathit{P}_{1}^{\left( \alpha
_{-3},\beta _{-3}\right) }\left( z\right) ,\mathit{P}_{2}^{\left( \alpha
_{-3},\beta _{-3}\right) }\left( z\right) ,\mathit{P}_{4}^{\left( \alpha
_{-3},\beta _{-3}\right) }\left( z\right) ,\mathit{P}_{5}^{\left( \alpha
_{-3},\beta _{-3}\right) }\left( z\right) \mid z\right) },  \notag
\end{eqnarray}

We have \cite{szego,magnus}

\begin{equation}
\frac{d^{k}}{dz^{k}}\left( \mathit{P}_{n}^{\left( \alpha ,\beta \right)
}\left( z\right) \right) =\frac{(1+\alpha +\beta +n)_{k}}{2^{k}}\mathit{P}%
_{n-k}^{\left( \alpha _{k},\beta _{k}\right) }\left( z\right)
\end{equation}%
and from the Rodrigues formula for the Jacobi polynomials \cite{magnus,szego}%
, it results

\begin{equation}
\frac{d^{k}}{dz^{k}}\left( \left( 1-z\right) ^{-\alpha }\left( 1+z\right)
^{-\beta }\mathit{P}_{n}^{\left( -\alpha ,-\beta \right) }\left( z\right)
\right) =\left( -2\right) ^{k}\frac{\left( n+k\right) !}{n!}\left(
1-z\right) ^{-\alpha _{k}}\left( 1+z\right) ^{-\beta _{k}}\mathit{P}%
_{n+k}^{\left( -\alpha _{k},-\beta _{k}\right) }\left( z\right) .
\end{equation}

Eq(\ref{wronskjacobi}) then gives the following bilinear determinantal
identity for Jacobi polynomials

\begin{equation}
\frac{\Pi _{-3,1,2,j}(z)}{\Pi _{-3,1,2}(z)}\sim \frac{\Pi _{1,2,4,5,j+3}(z)}{%
\Pi _{1,2,4,5}(z)},
\end{equation}%
where

\begin{equation}
\small
\Pi _{-3,1,2,j}(z)=\left\vert 
\begin{array}{cccc}
\left( 1-z^{2}\right) ^{3}\mathit{P}_{2}^{\left( -\alpha ,-\beta \right)
}\left( z\right) & \mathit{P}_{1}^{\left( \alpha ,\beta \right) }\left(
z\right) & \mathit{P}_{2}^{\left( \alpha ,\beta \right) }\left( z\right) & 
\mathit{P}_{j}^{\left( \alpha ,\beta \right) }\left( z\right) \\ 
\left( -2\right) \left( n+1\right) \left( 1-z^{2}\right) ^{2}\mathit{P}%
_{3}^{\left( -\alpha _{1},-\beta _{1}\right) }\left( z\right) & \frac{%
(2+\alpha +\beta )_{1}}{2}\mathit{P}_{0}^{\left( \alpha _{1},\beta
_{1}\right) }\left( z\right) & \frac{(3+\alpha +\beta )_{1}}{2}\mathit{P}%
_{1}^{\left( \alpha _{1},\beta _{1}\right) }\left( z\right) & \frac{%
(1+\alpha +\beta +j)_{1}}{2}\mathit{P}_{j-1}^{\left( \alpha _{1},\beta
_{1}\right) }\left( z\right) \\ 
4\frac{\left( n+2\right) !}{n!}\left( 1-z^{2}\right) \mathit{P}_{4}^{\left(
-\alpha _{2},-\beta _{2}\right) }\left( z\right) & 0 & \frac{(3+\alpha
+\beta )_{2}}{4}\mathit{P}_{0}^{\left( \alpha _{2},\beta _{2}\right) }\left(
z\right) & \frac{(1+\alpha +\beta +j)_{2}}{4}\mathit{P}_{j-2}^{\left( \alpha
_{2},\beta _{2}\right) }\left( z\right) \\ 
\left( -8\right) \frac{\left( n+3\right) !}{n!}\mathit{P}_{5}^{\left(
-\alpha _{3},-\beta _{3}\right) }\left( z\right) & 0 & 0 & \frac{(1+\alpha
+\beta +j)_{3}}{8}\mathit{P}_{j-3}^{\left( \alpha _{3},\beta _{3}\right)
}\left( z\right)%
\end{array}%
\right\vert ,
\end{equation}

\begin{equation}
\small
\Pi _{-3,1,2}(z)=\left\vert 
\begin{array}{ccc}
\left( 1-z^{2}\right) \mathit{P}_{2}^{\left( -\alpha ,-\beta \right) }\left(
z\right) & \mathit{P}_{1}^{\left( \alpha ,\beta \right) }\left( z\right) & 
\mathit{P}_{2}^{\left( \alpha ,\beta \right) }\left( z\right) \\ 
\left( -2\right) \left( n+1\right) \left( 1-z^{2}\right) \mathit{P}%
_{3}^{\left( -\alpha _{1},-\beta _{1}\right) }\left( z\right) & \frac{%
(2+\alpha +\beta )_{1}}{2}\mathit{P}_{0}^{\left( \alpha _{1},\beta
_{1}\right) }\left( z\right) & \frac{(3+\alpha +\beta )_{1}}{2}\mathit{P}%
_{1}^{\left( \alpha _{1},\beta _{1}\right) }\left( z\right) \\ 
4\frac{\left( n+2\right) !}{n!}\mathit{P}_{4}^{\left( -\alpha _{2},-\beta
_{2}\right) }\left( z\right) & 0 & \frac{(3+\alpha +\beta )_{2}}{4}\mathit{P}%
_{0}^{\left( \alpha _{2},\beta _{2}\right) }\left( z\right)%
\end{array}%
\right\vert ,
\end{equation}

\begin{equation}
\small
\Pi _{1,2,4,5}(z)=\left\vert 
\begin{array}{cccc}
\mathit{P}_{1}^{\left( \alpha _{-3},\beta _{-3}\right) }\left( z\right) & 
\mathit{P}_{2}^{\left( \alpha _{-3},\beta _{-3}\right) }\left( z\right) & 
\mathit{P}_{4}^{\left( \alpha _{-3},\beta _{-3}\right) }\left( z\right) & 
\mathit{P}_{5}^{\left( \alpha _{-3},\beta _{-3}\right) }\left( z\right) \\ 
\frac{(\alpha +\beta -4)_{1}}{2}\mathit{P}_{0}^{\left( \alpha _{-2},\beta
_{-2}\right) }\left( z\right) & \frac{(\alpha +\beta -3)_{1}}{2}\mathit{P}%
_{1}^{\left( \alpha _{-2},\beta _{-2}\right) }\left( z\right) & \frac{%
(\alpha +\beta -1)_{1}}{2}\mathit{P}_{3}^{\left( \alpha _{-2},\beta
_{-2}\right) }\left( z\right) & \frac{(\alpha +\beta )_{1}}{2}\mathit{P}%
_{4}^{\left( \alpha _{-2},\beta _{-2}\right) }\left( z\right) \\ 
0 & \frac{(\alpha +\beta -3)_{2}}{4}\mathit{P}_{0}^{\left( \alpha
_{-1},\beta _{-1}\right) }\left( z\right) & \frac{(\alpha +\beta -1)_{2}}{4}%
\mathit{P}_{2}^{\left( \alpha _{-1},\beta _{-1}\right) }\left( z\right) & 
\frac{(\alpha +\beta )_{2}}{4}\mathit{P}_{3}^{\left( \alpha _{-1},\beta
_{-1}\right) }\left( z\right) \\ 
0 & 0 & \frac{(\alpha +\beta -1)_{3}}{8}\mathit{P}_{1}^{\left( \alpha ,\beta
\right) }\left( z\right) & \frac{(\alpha +\beta )_{3}}{8}\mathit{P}%
_{2}^{\left( \alpha ,\beta \right) }\left( z\right) \\ 
0 & 0 & \frac{(\alpha +\beta -1)_{4}}{16}\mathit{P}_{0}^{\left( \alpha
_{1},\beta _{1}\right) }\left( z\right) & \frac{(\alpha +\beta )_{4}}{16}%
\mathit{P}_{1}^{\left( \alpha _{1},\beta _{1}\right) }\left( z\right)%
\end{array}%
\right\vert
\end{equation}%
and $\Pi _{1,2,4,5,j+3}(z)$ is equal to

\begin{equation}
\scriptsize
\left\vert 
\begin{array}{ccccc}
\mathit{P}_{1}^{\left( \alpha _{-3},\beta _{-3}\right) }\left( z\right)  & 
\mathit{P}_{2}^{\left( \alpha _{-3},\beta _{-3}\right) }\left( z\right)  & 
\mathit{P}_{4}^{\left( \alpha _{-3},\beta _{-3}\right) }\left( z\right)  & 
\mathit{P}_{5}^{\left( \alpha _{-3},\beta _{-3}\right) }\left( z\right)  & 
\mathit{P}_{j+3}^{\left( \alpha _{-3},\beta _{-3}\right) }\left( z\right) 
\\ 
\frac{(\alpha +\beta -4)_{1}}{2}\mathit{P}_{0}^{\left( \alpha _{-2},\beta
_{-2}\right) }\left( z\right)  & \frac{(\alpha +\beta -3)_{1}}{2}\mathit{P}%
_{1}^{\left( \alpha _{-2},\beta _{-2}\right) }\left( z\right)  & \frac{%
(\alpha +\beta -1)_{1}}{2}\mathit{P}_{3}^{\left( \alpha _{-2},\beta
_{-2}\right) }\left( z\right)  & \frac{(\alpha +\beta )_{1}}{2}\mathit{P}%
_{4}^{\left( \alpha _{-2},\beta _{-2}\right) }\left( z\right)  & \frac{%
(\alpha +\beta +j-2)_{1}}{2}\mathit{P}_{j+2}^{\left( \alpha _{-2},\beta
_{-2}\right) }\left( z\right)  \\ 
0 & \frac{(\alpha +\beta -3)_{2}}{4}\mathit{P}_{0}^{\left( \alpha
_{-1},\beta _{-1}\right) }\left( z\right)  & \frac{(\alpha +\beta -1)_{2}}{4}%
\mathit{P}_{2}^{\left( \alpha _{-1},\beta _{-1}\right) }\left( z\right)  & 
\frac{(\alpha +\beta )_{2}}{4}\mathit{P}_{3}^{\left( \alpha _{-1},\beta
_{-1}\right) }\left( z\right)  & \frac{(\alpha +\beta +j-2)_{2}}{4}\mathit{P}%
_{j+1}^{\left( \alpha _{-1},\beta _{-1}\right) }\left( z\right)  \\ 
0 & 0 & \frac{(\alpha +\beta -1)_{3}}{8}\mathit{P}_{1}^{\left( \alpha ,\beta
\right) }\left( z\right)  & \frac{(\alpha +\beta )_{3}}{8}\mathit{P}%
_{2}^{\left( \alpha ,\beta \right) }\left( z\right)  & \frac{(\alpha +\beta
+j-2)_{3}}{8}\mathit{P}_{j}^{\left( \alpha ,\beta \right) }\left( z\right) 
\\ 
0 & 0 & \frac{(\alpha +\beta -1)_{4}}{16}\mathit{P}_{0}^{\left( \alpha
_{1},\beta _{1}\right) }\left( z\right)  & \frac{(\alpha +\beta )_{4}}{16}%
\mathit{P}_{1}^{\left( \alpha _{1},\beta _{1}\right) }\left( z\right)  & 
\frac{(\alpha +\beta +j-2)_{4}}{16}\mathit{P}_{j-1}^{\left( \alpha
_{1},\beta _{1}\right) }\left( z\right)  \\ 
0 & 0 & 0 & \frac{(\alpha +\beta )_{5}}{32}\mathit{P}_{0}^{\left( \alpha
_{2},\beta _{2}\right) }\left( z\right)  & \frac{(\alpha +\beta +j-2)_{5}}{32%
}\mathit{P}_{j-2}^{\left( \alpha _{2},\beta _{2}\right) }\left( z\right) 
\end{array}%
\right\vert .
\end{equation}

\subsection{The Morse potential}

The Morse potential is the second exceptional primary TSIP of the first
category \cite{grandati}. It is defined as \cite{cooper,Dutt}

\begin{equation}
V(y;\omega ,\alpha )=\omega ^{2}y^{2}-2\alpha \omega y+\left( \alpha
-1/2\right) ^{2},\ \omega \in \mathbb{R},\left\vert \alpha \right\vert >1/2,
\label{potMorse}
\end{equation}%
where $y=\exp \left( -x\right) >0,\ x\in \mathbb{R}$. Contrarily to the
preceding examples, the Morse potential possesses only a finite number of
bound states. If we choose $\omega >0,\alpha >1/2$, it has exactly $\left[
\alpha -1/2\right] $ bound states ( $\left[ a\right] $ being the integer
part of $a$) which are given by

\begin{equation}
\psi _{n}\left( x;\omega ,\alpha \right) \sim y^{\alpha _{n}-1/2}e^{-\omega
y}\mathit{L}_{n}^{2\alpha _{n}-1}(2\omega y),\ n\in \left\{ 0,...,\left[
\alpha -3/2\right] \right\} ,  \label{foMorse}
\end{equation}%
with the corresponding energies $E_{n}(\alpha )=\left( \alpha -1/2\right)
^{2}-\left( \alpha _{n}-1/2\right) ^{2}=n\left( 2\alpha -1-n\right) ,$ where 
$\alpha _{k}=\alpha -k$.

The only parameter transformation under which the Morse potential Eq(\ref%
{potMorse}) is covariant, is the $\Gamma _{3}$ symmetry given by

\begin{equation}
\left( \omega ,\alpha \right) \overset{\Gamma _{3}}{\rightarrow }\left(
-\omega ,-\alpha \right) ,\left\{ 
\begin{array}{c}
V(x;\omega ,\alpha )\overset{\Gamma _{3}}{\rightarrow }V(x;-\omega ,-\alpha
)-E_{-1}\left( \alpha \right) \\ 
\psi _{n}(x;\omega ,\alpha )\overset{\Gamma _{3}}{\rightarrow }\psi
_{n}(x;-\omega ,-\alpha ).%
\end{array}%
\right.  \label{discreteMorse}
\end{equation}

Applying the symmetry $\Gamma _{3}$ on the Schr\"{o}dinger equation for $%
\psi _{n}$, we obtain

\begin{equation}
\widehat{H}(\omega ,\alpha )\psi _{n}(x;-\omega ,-\alpha )=\left(
E_{n}\left( -\alpha \right) +E_{-1}\left( \alpha \right) \right) \psi
_{n}(x;-\omega ,-\alpha )=E_{-(n+1)}\left( \alpha \right) \psi
_{n}(x;-\omega ,-\alpha ),
\end{equation}%
where $E_{-(n+1)}\left( \alpha \right) =-\left( n+1\right) \left( 2\alpha
+n\right) <0$, even if $n>\left[ \alpha -3/2\right] $. This implies ($\left(
-\alpha \right) _{n}=-\alpha _{-n}=-\left( \alpha +n\right) $)

\begin{equation}
\psi _{-(n+1)}(x;\omega ,\alpha )=\psi _{n}(x;-\omega ,-\alpha )=y^{-\alpha
_{-n}-1/2}e^{\omega y}\mathit{L}_{n}^{-2\alpha _{-n}-1}(-2\omega y),\ n\geq
0.
\end{equation}

Note that if $\alpha >1/2$, we have $\alpha _{-n}>1/2$ without any
restriction on $n$. Using Eq(\ref{asymptLag}), we obtain

\begin{equation}
\left\{ 
\begin{array}{c}
\psi _{-(n+1)}(x;\omega ,\alpha )\underset{x\rightarrow -\infty
,y\rightarrow +\infty }{\sim }y^{n-\alpha _{-n}-1/2}e^{\omega y}\rightarrow
+\infty \\ 
\psi _{-(n+1)}(x;\omega ,\alpha )\underset{x\rightarrow +\infty
,y\rightarrow 0^{+}}{\rightarrow }\left( -1\right) ^{n}y^{-\left( \alpha
+n+1/2\right) }\left( 2\alpha +2n\right) ...\left( 2\alpha +n+1\right)
\rightarrow \pm \infty .%
\end{array}%
\right.  \label{asymptMorse}
\end{equation}%
with $\pm =\left( -1\right) ^{n}$. $\psi _{-(n+1)}$ being in the
disconjugacy sector of $V(x;\omega ,\alpha )$, the disconjugacy theorem
ensures that for every odd value of $n\geq 0$, $\psi _{-n}$\ has no node on $%
\mathbb{R}
$. Consequently $A\left( w_{-(n+1)}\right) $ is a regular state-adding DBT
and $V^{\left( -n\right) }(x;\omega ,\alpha )$ is a regular extension of $%
V(x;\omega ,\alpha )$. As in the case of the harmonic oscillator, there is
no limitation on the possible values of $n$ since $\alpha _{-n}=\alpha
+n>1/2 $ for every positive $n$. Moreover, even when $n>\left[ \alpha -3/2%
\right] $, $\psi _{-(n+1)}$ stays an adequate seed function for a
state-adding DBT. This is in agreement with the fact that for every odd $n$
the one-step extension $V^{\left( -n\right) }$, entering then in the frame
of the extended Krein-Adler theorem, has an even gap in its spectrum. The
chain of complete state-adding DBT we can consider are then not limited.\ 

From Theorem \ref{thm1}, we can write

\begin{equation}
V^{\left( -M//n_{1},...,n_{k}\right) }(x;\omega ,\alpha )=V^{\left(
n_{1}+m,...,n_{k}+m\right) }(x;\omega ,\alpha _{-m})-\left( m+1\right)
\left( 2\alpha +m\right) .
\end{equation}%
the eigenstates of $V^{\left( -M//n_{1},...,n_{k}\right) }$ statisfying (see
Theorem \ref{thm1}) for $-m\leq j\leq \left[ \alpha -3/2\right] -m$

\begin{eqnarray}
\psi _{j}^{\left( -M//n_{1},...,n_{k}\right) }(x;\omega ,\alpha ) &=&\frac{%
W^{\left( -M//n_{1},...,n_{k,}j\right) }(x;\omega ,\alpha )}{W^{\left(
-M//n_{1},...,n_{k}\right) }(x;\omega ,\alpha )} \\
&\sim &\frac{W^{\left( n_{1}+m,...,n_{k}+m,j+m\right) }\left( x;\omega
,\alpha _{-m}\right) }{W^{\left( n_{1}+m,...,n_{k}+m\right) }\left( x;\omega
,\alpha _{-m}\right) }=\psi _{j+m}^{\left( n_{1}+m,...,n_{k}+m\right)
}(x;\omega ,\alpha _{-m}).\quad   \notag
\end{eqnarray}

In the same example as before (see Eq(\ref{Th1Appl}))

\begin{equation}
V^{\left( -3,1,2\right) }(x;\omega ,\alpha )=V^{\left( 1,2,4,5\right)
}(x;\omega ,\alpha )-4\left( 2\alpha +3\right)
\end{equation}%
and

\begin{equation}
\frac{W\left( \psi _{-3},\psi _{1},\psi _{2},\psi _{j}\mid x;\omega ,\alpha
\right) }{W\left( \psi _{-3},\psi _{1},\psi _{2}\mid x;\omega ,\alpha
\right) }\sim \frac{W\left( \psi _{1},\psi _{2},\psi _{4},\psi _{5},\psi
_{j+3}\mid x;\omega ,\alpha _{-3}\right) }{W\left( \psi _{1},\psi _{2},\psi
_{4},\psi _{5}\mid x;\omega ,\alpha _{-3}\right) }.
\end{equation}

Using Eq(\ref{spec OI}) and Eq(\ref{L3Lag}) and usual properties of
Wronskians \cite{muir}, we obtain ($z=2\omega y$)

\begin{eqnarray}
&&\frac{W\left( z^{-2\alpha -2}e^{z}\mathit{L}_{2}^{-2\alpha -3}\left(
-z\right) ,z^{-1}\mathit{L}_{1}^{2\alpha -3}\left( z\right) ,z^{-2}\mathit{L}%
_{2}^{2\alpha -5}\left( z\right) ,z^{-j}\mathit{L}_{j}^{2\alpha -2j-1}\left(
z\right) \mid x\right) }{W\left( z^{-2\alpha -2}e^{z}\mathit{L}%
_{2}^{-2\alpha -3}\left( -z\right) ,z^{-1}\mathit{L}_{1}^{2\alpha -3}\left(
z\right) ,z^{-2}\mathit{L}_{2}^{2\alpha -5}\left( z\right) \mid x\right) } \\
&\sim &\frac{W\left( z^{2}\mathit{L}_{1}^{2\alpha +3}\left( z\right) ,z%
\mathit{L}_{2}^{2\alpha +1}\left( z\right) ,z^{-1}\mathit{L}_{4}^{2\alpha
-3}\left( z\right) ,z^{-2}\mathit{L}_{5}^{2\alpha -5}\left( z\right) ,z^{-j}%
\mathit{L}_{j+3}^{2\alpha -2j-1}\left( z\right) \mid x\right) }{W\left( z^{2}%
\mathit{L}_{1}^{2\alpha +3}\left( z\right) ,z\mathit{L}_{2}^{2\alpha
+1}\left( z\right) ,z^{-1}\mathit{L}_{4}^{2\alpha -3}\left( z\right) ,z^{-2}%
\mathit{L}_{5}^{2\alpha -5}\left( z\right) \mid x\right) },  \notag
\end{eqnarray}%
which, with $W\left( y_{1},...,y_{m}\mid x\right) =\left( \frac{dz}{dx}%
\right) ^{m(m-1)/2}W\left( y_{1},...,y_{m}\mid z\right) $ \cite{muir}, leads
to the following bilinear Wronskian identity

\begin{eqnarray}
&&\frac{W\left( z^{-2\alpha -2}e^{z}\mathit{L}_{2}^{-2\alpha -3}\left(
-z\right) ,z^{-1}\mathit{L}_{1}^{2\alpha -3}\left( z\right) ,z^{-2}\mathit{L}%
_{2}^{2\alpha -5}\left( z\right) ,z^{-j}\mathit{L}_{j}^{2\alpha -2j-1}\left(
z\right) \mid z\right) }{W\left( z^{-2\alpha -2}e^{z}\mathit{L}%
_{2}^{-2\alpha -3}\left( -z\right) ,z^{-1}\mathit{L}_{1}^{2\alpha -3}\left(
z\right) ,z^{-2}\mathit{L}_{2}^{2\alpha -5}\left( z\right) \mid z\right) }
\label{wronsklag2} \\
&\sim &z\frac{W\left( z^{2}\mathit{L}_{1}^{2\alpha +3}\left( z\right) ,z%
\mathit{L}_{2}^{2\alpha +1}\left( z\right) ,z^{-1}\mathit{L}_{4}^{2\alpha
-3}\left( z\right) ,z^{-2}\mathit{L}_{5}^{2\alpha -5}\left( z\right) ,z^{-j}%
\mathit{L}_{j+3}^{2\alpha -2j-1}\left( z\right) \mid z\right) }{W\left( z^{2}%
\mathit{L}_{1}^{2\alpha +3}\left( z\right) ,z\mathit{L}_{2}^{2\alpha
+1}\left( z\right) ,z^{-1}\mathit{L}_{4}^{2\alpha -3}\left( z\right) ,z^{-2}%
\mathit{L}_{5}^{2\alpha -5}\left( z\right) \mid z\right) }.  \notag
\end{eqnarray}

\subsection{The effective radial Kepler-Coulomb}

The effective radial Kepler-Coulomb (ERKC) potential with zero ground level (%
$E_{0}(\alpha )=0$) is the third and last exceptional primary TSIP of the
first category \cite{grandati}. It is defined on the positive half line $x>0$
as%
\begin{equation}
V(x;\alpha )=\frac{\left( \alpha +1/2\right) (\alpha -1/2)}{x^{2}}-\frac{%
\gamma }{x}+\frac{\gamma ^{2}}{4\left( \alpha +1/2\right) ^{2}},\ \gamma
>0,\ \left\vert \alpha \right\vert >1/2.  \label{potKC}
\end{equation}

Choosing $\alpha >1/2$, its bound states are given by ($z_{n}=\gamma
x/\left( \alpha _{n}+1/2\right) $)

\begin{equation}
\psi _{n}\left( x;\alpha \right) =x^{\alpha +1/2}e^{-z_{n}/2}\mathit{L}%
_{n}^{2\alpha }(z_{n}),\ n\geq 0,  \label{foKC}
\end{equation}%
with the corresponding energies 
\begin{equation}
E_{n}(\alpha )=\frac{\gamma ^{2}}{4\left( \alpha +1/2\right) ^{2}}-\frac{%
\gamma ^{2}}{4\left( \alpha _{n}+1/2\right) ^{2}},
\end{equation}%
where $\alpha _{k}=\alpha +k.$

The only covariance transformation for the ERKC potentials is the $\Gamma
_{3}$ given by

\begin{equation}
\alpha \overset{\Gamma _{3}}{\rightarrow }\left( -\alpha \right) ,\left\{ 
\begin{array}{c}
V(x;\alpha )\overset{\Gamma _{3}}{\rightarrow }V(x;\alpha )-E_{-1}\left(
\alpha \right) \\ 
\psi _{n}(x;\alpha )\overset{\Gamma _{3}}{\rightarrow }\psi _{n}(x;-\alpha ),%
\end{array}%
\right.  \label{gamma3KC}
\end{equation}%
with 
\begin{equation}
\alpha _{k}\overset{\Gamma _{3}}{\rightarrow }-\alpha +k=-\alpha _{-k},\quad
E_{n}\left( \alpha \right) \overset{\Gamma _{3}}{\rightarrow }\gamma
^{2}/4\left( \frac{1}{\left( \alpha _{-1}+1/2\right) ^{2}}-\frac{1}{\left(
\alpha _{-\left( n+1\right) }+1/2\right) ^{2}}\right) =E_{-(n+1)}\left(
\alpha \right) -E_{-1}\left( \alpha \right) .
\end{equation}

We then have

\begin{equation}
\widehat{H}(x;\alpha )\psi _{n}(x;-\alpha )=E_{-(n+1)}\left( \alpha \right)
\psi _{n}(x;-\alpha ),
\end{equation}%
which leads to

\begin{equation}
\psi _{-\left( n+1\right) }(x;\alpha )=\psi _{n}(x;-\alpha )=x^{-\alpha
+1/2}\exp \left( z_{-\left( n+1\right) }/2\right) \mathit{L}_{n}^{-2\alpha
}(-z_{-\left( n+1\right) }).  \label{phi3KC}
\end{equation}

Since we have $E_{-(n+1)}\left( \alpha \right) =1/\left( \alpha +1/2\right)
^{2}-1/\left( \alpha _{-\left( n+1\right) }+1/2\right) ^{2}<0$ for every
value of $n\geq 0$, $\psi _{-\left( n+1\right) }(x;\alpha )$ is in the
disconjugacy sector of $V(x;\alpha )$. Using Eq(\ref{asymptLag}) and the
disconjugacy theorem, we deduce that, if $\alpha >n+1/2$ 
\begin{equation}
\left\{ 
\begin{array}{c}
\psi _{-(n+1)}(x;\alpha )\underset{x\rightarrow 0^{+}}{\sim }x^{-\alpha +1/2}%
\frac{\left( -2\right) ^{n}}{n!}\left( \alpha -1/2\right) ...\left( \alpha
-n\right) \rightarrow \pm \infty \\ 
\psi _{-(n+1)}(x;\alpha )\underset{x\rightarrow +\infty }{\sim }\frac{%
x^{n-\alpha +1/2}}{\left( \alpha -n-1/2\right) ^{n}n!}\exp \left( \frac{%
\gamma x}{2\left( \alpha -n-1/2\right) }\right) \rightarrow +\infty ,%
\end{array}%
\right.  \label{asymptERKC}
\end{equation}%
with $\pm =\left( -1\right) ^{n}$. Consequently, $\psi _{-n}$ is free of
node for node for $n$ odd and $A(w_{-n})$ is then a state-adding DBT. If $%
1/2<\alpha <n+1/2$

\begin{equation}
\left\{ 
\begin{array}{c}
\psi _{-(n+1)}(x;\alpha )\underset{x\rightarrow 0^{+}}{\sim }x^{-\alpha +1/2}%
\frac{2^{n}}{n!}\left( \alpha -1/2\right) ...\left( \alpha -n\right)
\rightarrow \pm \infty \\ 
\psi _{-(n+1)}(x;\alpha )\underset{x\rightarrow +\infty }{\sim }\frac{%
x^{n-\alpha +1/2}}{\left( \alpha -n-1/2\right) ^{n}n!}\exp \left( \frac{%
\gamma x}{2\left( \alpha -n-1/2\right) }\right) \rightarrow 0^{\pm },%
\end{array}%
\right.  \label{asymptERKC2}
\end{equation}%
and even if $A(w_{-n})$ is regular, it is not state-adding but $V^{\left(
-n\right) }$ is strictly isospectral to $V$, as already noted in \cite%
{grandati} by using the Kienast-Lawton-Hahn theorem. Note that the
constraint is coherent with the fact that to enter into the frame of the
extended Krein-Adler theorem, we have, as in the case of the isotonic
oscillator, to ensure the regularity of all the extensions $V^{\left(
-j\right) }$ of the complete chain $\left( -M\right) $, namely the condition 
$\left\vert \alpha _{-j}\right\vert <1/2,\ \forall j\leq m$.

If $\alpha >m+1/2$, the extended potential $V^{\left(
-M//n_{1},...,n_{k}\right) }$ is regular and satisfies (see Theorem \ref{thm1})

\begin{equation}
V^{\left( -M//n_{1},...,n_{k}\right) }(x;\alpha )=V^{\left(
n_{1}+m,...,n_{k}+m\right) }(x;\alpha _{-m})-2m\omega ,
\end{equation}%
with eigenstates ($j\geq -m$)

\begin{equation}
\psi _{j}^{\left( -M//n_{1},...,n_{k}\right) }(x;\alpha )=\frac{W^{\left(
-M//n_{1},...,n_{k,}j\right) }(x;\alpha )}{W^{\left(
-M//n_{1},...,n_{k}\right) }(x;\alpha )}\sim \frac{W^{\left(
n_{1}+m,...,n_{k}+m,j+m\right) }\left( x;\alpha _{-m}\right) }{W^{\left(
n_{1}+m,...,n_{k}+m\right) }\left( x;\alpha _{-m}\right) }=\psi
_{j+m}^{\left( n_{1}+m,...,n_{k}+m\right) }(x;\alpha _{-m}).\quad
\end{equation}

For instance, (see Eq(\ref{Th1Appl})) for $\alpha >7/2$

\begin{equation}
V^{\left( -3,1,2\right) }(x;\alpha )=V^{\left( 1,2,4,5\right) }(x;\alpha
)-3\omega .
\end{equation}%
and

\begin{equation}
\frac{W\left( \psi _{-3},\psi _{1},\psi _{2},\psi _{j}\mid x;\omega ,\alpha
\right) }{W\left( \psi _{-3},\psi _{1},\psi _{2}\mid x;\omega ,\alpha
\right) }\sim \frac{W\left( \psi _{1},\psi _{2},\psi _{4},\psi _{5},\psi
_{j+3}\mid x;\omega ,\alpha _{-3}\right) }{W\left( \psi _{1},\psi _{2},\psi
_{4},\psi _{5}\mid x;\omega ,\alpha _{-3}\right) }.
\end{equation}

Using Eq(\ref{foKC}) and Eq(\ref{phi3KC}) and usual properties of Wronskians 
\cite{muir}, this leads to the following bilinear Wronskian identity

\begin{eqnarray}
&&\frac{W\left( x^{-2\alpha }e^{z_{-3}/2}\mathit{L}_{2}^{-2\alpha }\left(
-z_{-3}\right) ,e^{-z_{1}/2}\mathit{L}_{1}^{2\alpha }\left( z_{1}\right)
,e^{-z_{2}/2}\mathit{L}_{2}^{2\alpha }\left( z\right) ,e^{-z_{j}/2}\mathit{L}%
_{j}^{2\alpha }\left( z\right) \mid x\right) }{W\left( x^{-2\alpha
}e^{z_{-2}/2}\mathit{L}_{2}^{-2\alpha }\left( -z_{-2}\right) ,e^{-z_{1}/2}%
\mathit{L}_{1}^{2\alpha }\left( z_{1}\right) ,e^{-z_{2}/2}\mathit{L}%
_{2}^{2\alpha }\left( z\right) \mid x\right) } \\
&\sim &\frac{W\left( e^{-z_{-2}/2}\mathit{L}_{1}^{\alpha _{-3}}\left(
z_{-2}\right) ,e^{-z_{-1}/2}\mathit{L}_{2}^{\alpha _{-3}}\left(
z_{-1}\right) ,e^{-z_{1}/2}\mathit{L}_{4}^{\alpha _{-3}}\left( z_{1}\right)
,e^{-z_{2}/2}\mathit{L}_{5}^{\alpha _{-3}}\left( z_{2}\right) ,e^{-z_{j}/2}%
\mathit{L}_{j+3}^{\alpha _{-3}}\left( z_{j}\right) \mid x\right) }{W\left(
e^{-z_{-2}/2}\mathit{L}_{1}^{\alpha _{-3}}\left( z_{-2}\right) ,e^{-z_{-1}/2}%
\mathit{L}_{2}^{\alpha _{-3}}\left( z_{-1}\right) ,e^{-z_{1}/2}\mathit{L}%
_{4}^{\alpha _{-3}}\left( z_{1}\right) ,e^{-z_{2}/2}\mathit{L}_{5}^{\alpha
_{-3}}\left( z_{2}\right) \mid x\right) }.  \notag
\end{eqnarray}

\section{Conclusion}

We have shown that, when applied to a shape invariant potential, the Krein-Adler theorem admits an enlarged version which gives the regularity condition for
mixed chains of state-deleting and state-adding Darboux-B\"acklund transformations (DBTs). Moreover, the existence of a reverse shape invariance condition for complete chains of state-adding DBTs allows to establish determinantal bilinear identities for the classical orthogonal polynomials.

A more general version of this extended Krein-Adler theorem applicable to a
wider class of potentials will be addressed in a forthcoming work.


\end{document}